\documentclass[smallextended]{svjour3} 

\smartqed  

\usepackage{xcolor}

\usepackage{graphics}
\usepackage{graphicx}

\usepackage[english]{babel}
\usepackage{natbib}
\usepackage{url}

\usepackage{tabularx}
\usepackage{rotating}

\usepackage{xspace}

\def\RR{\textsf{R}\xspace}

\let\pkg=\strong
\newcommand\code{\bgroup\@codex}
\def\@codex#1{\small {\normalfont\ttfamily\hyphenchar\font=45 #1}\egroup}

\newcommand{\rholag}{\rho_{\scriptscriptstyle{\mathrm{Lag}}}}
\newcommand{\rhoerr}{\rho_{\scriptscriptstyle{\mathrm{Err}}}}

\newcommand{\logit}{\textrm{logit}}

\let\proglang=\textsf

\begin{document}


\title{Estimating Spatial Econometrics Models with Integrated Nested Laplace 
Approximation}

\author{Virgilio G\'omez-Rubio \and Roger S. Bivand \and H{\aa}vard Rue}

\institute{Virgilio G\'omez-Rubio \at
  Department of Mathematics, School of Industrial Engineering,
University of Castilla-La Mancha,
02071 Albacete, Spain.\\
\email{Virgilio.Gomez@uclm.es}
\and
Roger S. Bivand \at
Department of Economics,
Norwegian School of Economics,
Helleveien 30,
N-5045 Bergen, Norway.\\
\email{roger.bivand@nhh.no}
\and
H\r{a}vard Rue \at
CEMSE Division,
King Abdullah University of Science and Technology,
Thuwal 23955-6900, Saudi Arabia\\
\email{haavard.rue@kaust.edu.sa}
}

%
%

\date{Received: date / Accepted: date}

\maketitle

\begin{abstract}

The integrated Nested Laplace Approximation \citep{isi:000264374200002}
provides a fast and effective method for marginal inference on
Bayesian hierarchical models. This methodology has been implemented
in the \pkg{R-INLA} package which permits INLA to be used from within
\RR statistical software. Although INLA is implemented as a general
methodology, its use in practice is limited to the models implemented
in the \pkg{R-INLA} package.

Spatial autoregressive models are widely used in spatial econometrics but
have until now been missing from the \pkg{R-INLA} package.  In this paper,
we describe the implementation and application of a new class of latent
models in INLA made available through \pkg{R-INLA}. This new latent class
implements a standard spatial lag model.

The implementation of this latent model in \pkg{R-INLA} also means that
all the other features of INLA can be used for model fitting, model
selection and inference in spatial econometrics, as will be shown in this
paper. Finally, we will illustrate the use of this new latent model and
its applications with two datasets based on Gaussian and binary outcomes.

\end{abstract}







\section{Introduction}

Interest in the Bayesian analysis of regression models with spatial
dependence has existed since spatial econometrics came into being in
the late 1970s. \citet{hepple:79} and \citet{anselin:82} point to key
benefits, such as being able to make exact, finite-sample inferences
in models in which only large sample, asymptotic inferences would be
feasible, and in the examination of model robustness to specification
error \citep[][p. 180]{hepple:79}. \citet[][pp. 88--91]{Anselin:1988}
extends this discussion, but admits that Bayesian approaches had not at
that time been applied often in spatial econometrics. \citet{hepple:95a,
hepple:95b} continued to follow up topics within Bayesian estimation,
including Bayesian model choice \citep{hepple:04}. No software was
available until the Spatial Econometrics Library was made available
within the Econometrics Toolbox for Matlab, based on \citet{lesage:97,
lesage:00}.\footnote{See \url{http://www.spatial-econometrics.com/}.}

\citet{LeSagePace:2009} provide a summary of spatial econometrics models
and their applications. For Bayesian inference, they used Markov Chain
Monte Carlo (MCMC) algorithms to estimate the posterior distributions of
the model parameters. These techniques give a feasible way of fitting
Bayesian models, but can be computationally intensive. In addition,
while the Matlab MCMC implementation does provide user access to change
prior values from their defaults, the time required to check non-default
settings may be considerable.

\citet{Bivandetal:2014,Bivandetal:2015} describe how to use the Integrated
Nested Laplace Approximation \citep[INLA,][]{isi:000264374200002} to fit some
spatial econometrics models. They focus on some models based on spatial
autoregressive specifications on the response and the error terms often used in
spatial econometrics. Because of the lack of an implementation of these models
within the \pkg{R-INLA} software at that time,
\citet{Bivandetal:2014,Bivandetal:2015} fit many different models conditioning
on values of the spatial autocorrelation parameter. These conditional models
can be fitted with \pkg{R-INLA} and they are later combined using Bayesian
model averaging \citep[BMA,][]{Hoetingetal:1999} to obtain the posterior
marginals of the parameters of the desired model.
\cite{GomezRubioPalmiPerales:2019} discuss different approaches to fit spatial
econometrics with INLA and how to perform multivariate inference on the output.
Similarly, \cite{GomezRubioetal:2020} exploit Bayesian model averaging to fit
spatial econometrics models wiht INLA.

INLA is based on approximating the posterior marginal distributions of
the model parameters by means of different Laplace approximations. This
provides a numerically fast method to fit models that can be applied
to a wide range of research topics. INLA is restricted to models whose
latent effects are Gaussian Markov Random Fields, but this class of
models includes many models used in practice in a range of disciplines.

In this paper we describe the implementation of a new latent class, that
we will call \code{slm}, within \pkg{R-INLA} that facilitates fitting
spatial econometrics models. This provides an alternative to fitting
some of the models in the Spatial Econometrics Library.  In addition,
this creates a faster approach for Bayesian inference when only marginal
inference on the model parameters is required.

This new approach will make fitting a wide range of spatial econometrics models
very easy thought the \pkg{R-INLA} package. A flexible specification of these
models will allow the inclusion of smooth terms to explore non-linear
relationships between variables. The new latent effects for spatial econometrics
can be combined with other latent effects to fit more complex models.
Furthermore, models will be fitted faster than with traditional MCMC, so a
larger number of models can be explored and different model selection
techniques can be used.

This paper has the following structure. After providing background
descriptions of some spatial econometrics models and the Integrated Nested
Laplace Approximation, we introduce the new \code{slm} latent model in
Section \ref{sec:slm}. A summary on the use of different likelihoods is
included in Section \ref{sec:like}.  The computation of the impacts is
laid out in Section \ref{sec:impacts}. Section \ref{sec:appl} describes
some applications on model selection and section \ref{sec:other} deals
with other issues in Bayesian inference.  Examples are included in
Section \ref{sec:examples}, using the well-known Boston house price data
set and the Katrina business re-opening data set, and a final discussion
is given in Section \ref{sec:disc}.

\section{Background}

\subsection{Spatial Econometrics Models}
\label{sec:spateco}

In this section we summarise some of the spatial econometrics models
that we will use throughout this paper. For a review on spatial
econometrics models see \citet{Anselin:2010}.  We will follow the
notation used in \citet{Bivandetal:2014}, which is in turn  derived from
\citet{Anselin:1988} and \citet{LeSagePace:2009, lesage+pace10}.

We will assume that we have a vector $y$ of observations from $n$
different regions. The adjacency structure of these regions is available
in a matrix $W$, which may be defined in different ways. Unless otherwise
stated, we will use standard binary matrices to denote adjacency between
regions, with standardised rows.  This is helpful in offering known
bounds for the spatial autocorrelation parameters \citep[see,][for
details]{Haining:2003}.  Also, we will assume that the $p$ covariates
available are in a design matrix $X$, which will be used to construct
regression models. \citet{paceetal:12} have pointed out challenges
involved in estimating models of this kind that we intend to address in
further research.

The first model that we will describe is the Spatial Error Model (SEM)
which is based on a spatial autoregressive error term: 

\begin{equation}
y= X \beta+u; u=\rhoerr Wu+e; e\sim MVN(0, \sigma^2 I_n).
\end{equation}
\noindent
Here, $\rhoerr$ is the spatial autocorrelation parameter associated to the
error term. This measures how strong spatial dependence is.  $\beta$ is the
vector of coefficients of the covariates in the model.  The error term $e$ is
supposed to follow a multivariate Normal distribution with zero mean and
diagonal variance-covariance matrix $\sigma^2 I_n$.  $\sigma^2$ is a global
variance parameter while $I_n$ is the identity matrix of dimension $n\times
n$.

Alternatively, we can consider an autoregressive model on the response (Spatial Lag Model, SLM):

\begin{equation}
y = \rholag W y + X \beta + e; e\sim MVN(0, \sigma^2 I_n).
\end{equation}
\noindent
$\rholag$ is now the spatial autocorrelation parameter associated to the
autocorrelated term on the response. 

Next, a third model that is widely used in spatial econometrics is the
Spatial Durbin model (SDM):

\begin{equation}
y = \rholag W y + X \beta + W X \gamma + e;\ e\sim MVN(0, \sigma^2I_n).
\end{equation}
\noindent
$\gamma$ is a vector of coefficients for the spatially lagged covariates, shown as matrix $WX$.

A variation of this model is the Spatial Durbin Error Model (SDEM),
in which the error is autoregressive:

\begin{equation}
y = X \beta + W X \gamma + u;\ u=\rhoerr Mu+e;\ e\sim MVN(0, \sigma^2I_n).
\end{equation}
\noindent
Here $M$ is an adjacency matrix for the error term that may be different
from $W$.

All these models can be rewritten so that the response $y$ only appears
on the left hand side. The SEM model can also be written as

\begin{equation}
y= X \beta + (I_n-\rhoerr W)^{-1}e; e\sim MVN(0, \sigma^2 I_n);
\label{eq:sem}
\end{equation}
\noindent
the SLM model is equivalent to

\begin{equation}
y = (I_n-\rholag W)^{-1}(X\beta+e);\ e \sim MVN(0, \sigma^2I_n);
\label{eq:slm}
\end{equation}
\noindent
the SDM model is


\begin{equation}
y= (I_n-\rholag W)^{-1} (X^* \beta'+e);\ e\sim MVN(0, \sigma^2 I_n);
\label{eq:sdm}
\end{equation}
\noindent
with $X^*=[X, WX]$, the new matrix of covariates with the original and
the lagged covariates and $\beta'=[\beta, \gamma]$, the associated vector
of coefficients.
Finally, the SDEM can be written as
 
\begin{equation}
y= X^* \beta' + (I_n-\rhoerr M)^{-1} e;\ e\sim MVN(0, \sigma^2 I_n).
\label{eq:sdem}
\end{equation}

For completeness, we will include a simplified model without
spatial autocorrelation parameters and lagged variables (Spatially Lagged $X$ model, SLX):

\begin{equation}
y= X^* \beta' + e;\ e\sim MVN(0, \sigma^2 I_n),
\label{eq:slx}
\end{equation}

These are a set of standard models in spatial econometrics, focussing
on three key issues: spatially autocorrelated errors, spatially
autocorrelated responses and spatially lagged covariates. More complex
models can be built from these three standard models; the main difference
is that those models incorporate more than one spatial autocorrelation
parameter.

\subsection{The Integrated Nested Laplace Approximation} \label{sec:INLA}

Bayesian inference on hierarchical models has often relied on the use
of computational methods among which Markov Chain Monte Carlo is the
most widely used. In principle, MCMC has the advantage of being able
to handle a large number of models, but it has it drawbacks, such as
slow convergence of the Markov chains and the difficulty of obtaining
sampling distributions for complex models.

\citet{isi:000264374200002} have developed an approximate method to
estimate the marginal distributions of the parameters in a Bayesian
model. In particular, they focus on the family of Latent Gaussian Markov
Random Fields models. We describe here how this new methodology has been
developed, but we refer the reader to the original paper for details.

First of all, a vector of $n$ observed values $\mathbf{y} =
(y_1,\ldots,y_n)$ are assumed to be distributed according to
one of the distributions in the exponential family, with mean $\mu_i$.
Observed covariates and a linear predictor on them
(possibly plus random effects) may be linked to the mean $\mu_i$
by using an appropriate transformation (i.e., a link function). Hence,
this linear predictor $\eta_i$ may be made of a fixed term on the
covariates plus random effects and other non-linear terms.

The distribution of $\mathbf{y}$ will depend on a number of
hyperparameters $\theta_1$. The vector $\mathbf{x}$ of latent effects
forms a Gaussian Markov Random Field with precision matrix $Q(\theta_2)$,
where $\theta_2$ is a vector of hyperparameters. The hyperparameters
can be represented in a unique vector $\theta=(\theta_1, \theta_2)$. It
should be noted that observations $\mathbf{y}$ are independent given
the values of the latent effects $\mathbf{x}$ and the hyperparameters
$\theta$. This can be written as

\begin{equation}
\pi(\mathbf{y}|\mathbf{x},\theta) = \prod_{i\in\mathcal{I}} \pi(y_i|x_i,\theta)
\end{equation}
\noindent
Here, $x_i$ represents the linear predictor $\eta_i$ and $\mathcal{I}$ is
a vector of indices over $1,\ldots,n$. If there are missing values in
$\mathbf{y}$ these indices will not be included in $\mathcal{I}$.

The posterior distribution of the latent effects $\mathbf{x}$ and the
vector of hyperparameters $\theta$ can be written as

\begin{eqnarray}
\pi(\mathbf{x}, \mathbf{\theta}|\mathbf{y}) \propto
\pi(\mathbf{\theta}) \pi(\mathbf{x}|\mathbf{\theta})\prod_{i\in \mathcal{I}}\pi(y_i|x_i,\mathbf{\theta})
\propto \\\nonumber
\propto \pi(\mathbf{\theta}) |\mathbf{Q}(\mathbf{\theta})|^{1/2} \exp\{-\frac{1}{2}\mathbf{x}^T \mathbf{Q}(\mathbf{\theta}) \mathbf{x}+\sum_{i\in\mathcal{I}} \log(\pi(y_i|x_i, \mathbf{\theta}) \}.
\end{eqnarray}

INLA will not try to estimate the joint distribution $\pi(\mathbf{x},
\mathbf{\theta}|\mathbf{y})$ but the marginal distribution of single latent
effects and hyperparameters, i.e., $\pi(x_j|\mathbf{y})$ and
$\pi(\theta_k|\mathbf{y})$. Indices $j$ and $k$ will move in a different range
depending on the number of latent variables and hyperparameters.

INLA will first compute an approximation to $\pi(\theta|\bf{y})$, 
$\tilde\pi(\theta|\mathbf{y})$, that
will be used later to compute an approximation to $\pi(x_j|\mathbf{y})$. This
can be done because 

\begin{equation}
\pi(x_j|\mathbf{y}) = \int \pi(x_j|\mathbf{\theta}, \mathbf{y})  \pi(\mathbf{\theta}| \mathbf{y}) d\mathbf{\theta}.
\end{equation}
\noindent
Hence, an approximation can be developed as follows:

\begin{equation}
\tilde\pi(x_j|\mathbf{y})= 
\sum_g \tilde\pi (x_j|\mathbf{\theta}_g, \mathbf{y})\times 
\tilde\pi(\mathbf{\theta}_g|\mathbf{y})\times \Delta_g,
\end{equation}
\noindent
Here, $\mathbf{\theta}_g$ are values of the ensemble of hyperparameters in a
grid, with associated weights $\Delta_g$.
$\tilde\pi(x_j|\theta_g, \mathbf{y})$ is an approximation to $\pi
(x_j|\mathbf{\theta}_g, \mathbf{y})$ and this is thoroughly addressed in
\citet{isi:000264374200002}. They comment on the use of a Gaussian approximation
and others based on repeated Laplace Approximations and explore the error
of the approximation.

This methodology is implemented in the \pkg{R-INLA} package. It allows for an
easy access to many different types of likelihoods, latent models and priors
for model fitting.  However, this list is by no means exhaustive and there are
many latent effects that have not been implemented yet. This is the reason why
we describe a newly implemented \code{slm} latent effect that has many
applications in spatial econometrics.

\section{The \code{slm} latent model in \pkg{R-INLA}}
\label{sec:slm}

Although the INLA methodology covers a wide range of models, latent
models need to be implemented in compiled code in the INLA software to be
able to fit the models described earlier in this paper. Hence, the newly
implemented \code{slm} latent model fills the gap for spatial econometrics
models. This new latent model implements the following expression as a
random effect that can be included in the linear predictor:

\begin{equation}
\mathbf{x} = (I_n-\rho W)^{-1} (X\beta +\varepsilon)
\label{eq:slminla}
\end{equation}
\noindent
Here, 
$\mathbf{x}$
is a vector of $n$ random effects, $I_n$ is the identity matrix of
dimension $n\times n$, $\rho$ is a spatial autocorrelation parameter (that we
will discuss later), $W$ is a $n\times n$ weight matrix, $X$ a matrix of
covariates with coefficients $\beta$ and $\varepsilon$ is a vector of
independent Gaussian errors with zero mean and precision $\tau I_n$.

In this latent model, we need to assign  prior distributions to the vector of
coefficients $\beta$, spatial autocorrelation parameter $\rho$ and precision
of the error term $\tau$. By default,  $\beta$ takes a multivariate
Gaussian distribution with zero mean and precision matrix  $Q$
(which must be specified); $\logit(\rho)$ takes a Gaussian prior with
zero mean and precision 10; and, $\log(\tau)$ takes a log-gamma prior
with parameters 1 and $5\cdot10^{-5}$. Other priors can be assigned to these
hyperparameters following standard \pkg{R-INLA} procedures.

Note that, as described in Section \ref{sec:spateco}, spatial econometrics
models can be derived from this implementation. In particular, the SEM model
is a particular case of  equation (\ref{eq:slminla}) with $\beta=0$. The SLM
model can be fitted with no modification and the SDM model can be implemented
using a matrix of covariates made of the original covariates plus the lagged
covariates.

The SDEM model simply takes two terms, a standard linear term on the covariates
(and lagged covariates), plus a \code{slm} effect with $\beta=0$.  Finally, the
SLX model can be fitted using a standard linear regression on the covariates
and lagged covariates and typical i.i.d. random effects.

%
%
%

\subsection{Implementation}


We will describe here the implementation of the new \code{slm}
latent class. For a Gaussian response (and similarly for non-Gaussian
likelihoods) the model can be written as 

$$
y =  (I-\rho W)^{-1} (X\beta + \varepsilon) 
$$
\noindent
It can be rewritten using $\mathbf{x} = (I-\rho W)^{-1} (X\beta + \varepsilon)$, so that with 
observations $y$, then we have

$$
y = \mathbf{x} + e
$$
\noindent
where $e$ is a tiny error that is introduced to fit the model. This
is the error present in a Gaussian distribution and will not appear
if another likelihood is used.

By re-writing the \code{slm} as $\mathbf{x}$ in this way, we define it so that it suits the \code{f()}-component in the \pkg{R-INLA} framework. Given this, we note that the \code{slm} model is a Markov model with a sparse
precision matrix, and so conforms to the INLA framework. We provide
a detailed proof in \ref{app:GMRF} and show here the main
results.

The mean and precision of ($\mathbf{x}, \beta)$ given the hyperparameters
$\tau$ and $\rho$ are given by

\begin{equation}
E[\mathbf{x}, \beta|\tau, \rho] = 0
\label{eq:mujoint}
\end{equation}

\begin{equation}
Prec[\mathbf{x}, \beta|\tau, \rho] =  
\left[
\begin{array}{cc}
\tau (I_n-\rho W')(I_n-\rho W)   & -\tau (I_n-\rho W') X\\
-\tau X' (I_n-\rho W) & Q +\tau X'X\\
\end{array}
\right]
\label{eq:precjoint}
\end{equation}
\noindent
Note that this precision matrix is highly sparse and symmetric. Efficient
computation using this new latent effect can be carried out using the GMRF library, as described in \citet{RueHeld:2005}.

The full model can then be derived conditioning on different parameters.
Hence, the joint distribution of $\mathbf{x}$, $\beta$,  $\tau$  and $\rho$
can be written as

$$
\pi(\mathbf{x}, \beta, \tau, \rho) = \pi(\mathbf{x}, \beta| \tau, \rho) 
\pi(\tau, \rho)=
\pi(\mathbf{x}, \beta| \tau, \rho) \pi(\tau) \pi(\rho)
$$

$\pi(\mathbf{x}| \beta, \rho)$ is a Gaussian distribution with mean and
precision shown in equations (\ref{eq:mujoint}) and (\ref{eq:precjoint}),
respectively.  The prior distribution of $\beta$ is Gaussian with zero mean and
known precision matrix $Q$. This matrix is often taken diagonal to assume that
the coefficients are independent a priori.  Also, it may be worth rescaling the
covariates in $X$ to avoid numerical problems. Including lagged covariates may
lead to further numerical instability as they may be highly correlated with the
original covariates.

It is internally assumed that $\rho$ is between 0 and 1
so that a Gaussian prior is assigned to $\log(\rho/(1-\rho))$.
When computing $I_n-\rho W$, $\rho$ is  re-scaled to be in a range
of appropriate values. See details in the description of the \proglang{R}
interface in \ref{app:Rinterface}.

%
%
%


%

\section{Using different likelihoods}
\label{sec:like}

\subsection{Binary response}

The models described in Section \ref{sec:spateco}
assume a Gaussian response but other distributions can be used for the
response. \citet{LeSageetal:2011} consider a binary outcome when studying the
probability of re-opening a business in New Orleans in the aftermath of
Hurricane Katrina.  Binary outcome $y_i$ is modelled using a latent Gaussian
variable $y_i^*$ as follows:

\begin{equation}
y_i=
\left\{
\begin{array}{cc}
1 & \textrm{if}\ y_i^*\geq 0\\
0 & \textrm{if}\ y_i^*< 0\\
\end{array}
\right. .
\label{eq:probit}
\end{equation}
\noindent
$y_i^*$ is the net profit, so that if it is equal or higher than zero the
business will re-open. $y_i^*$ is assumed to be a Gaussian variable that can
be modelled using the spatial econometrics models described in Section
\ref{sec:spateco}. Note that in this case the variance of the error term is
set to 1 to avoid identifiability problems
\citep{LeSageetal:2011,Bivandetal:2014}.

Several authors have assessed different methods for the estimation of spatial
probit models.  \citet{Bille:2013} has compared the methods of Maximum
Likelihood (ML) and Generalised Method of Moments using a Monte Carlo study and
propose alternatives to avoid the inversion of $|I-\rho W|$ when fitting the
models.  \citet{CalabreseElkink:2014} compare a larger number of estimation
methods for spatial probit models, including MCMC algorithms, to estimate
the spatial autocorrelation parameter and provide a study on the predictive
quality of each method.


Instead of using a  ``broken-stick'' function such as the one shown in equation
(\ref{eq:probit}), \pkg{R-INLA} relies on standard logit and probit links, among
others. In our examples, the spatial probit is based on using a (continuous) probit link
function instead of the one shown in equation (\ref{eq:probit}). Hence, 
differences in some results can be expected.

\subsection{Other likelihoods}

\pkg{R-INLA} provides a number of likelihoods that can be used when
defining a model, so that an \code{slm} latent effect is included in
the linear predictor.  However, attention should be paid so that the
resulting model makes sense.  A particular problem of interest is
whether all parameters in the model are identifiable.  For example,
if the spatial probit model is used, $\tau$ must be set to one
so that $\beta$ can be properly estimated \citep{LeSagePace:2009}.
Using other highly parameterised likelihoods (such as, for example,
zero-inflated models) obliges the analyst to pay attention to details
to ensure that output makes sense.

\subsection{Additional  effects}
\label{subsec:addeff}

\pkg{R-INLA} makes it possible to include additional effects  in the linear
predictor.
All models presented so far assume a spatial structure on the error terms.
Like \citet{besagetal:1991}, it is possible to consider a model in
which there are two different random effects: one spatial with an
autocorrelated structured defined by the \code{slm} latent class plus
unstructured random effects. For example, the SLM model can be extended
as follows:

\begin{equation}
y = (I_n-\rholag W)^{-1} (X\beta+\epsilon)+u;\ u\sim MVN(0,\sigma^2 I_n)
\end{equation}
\noindent
Note that the random effect $u$ can have different structures.

Furthermore, other effects than linear can be explored for the covariates, as
\pkg{R-INLA} includes different types of smoothers, such as first and second order random walks.

\section{Computation of the Impacts}
\label{sec:impacts}

Impacts are related to how changes in the covariates in one
area will affect the response in other areas, and there are different types of
{\em impacts} to measure these effects.  As \citet[][Section
2.7]{LeSagePace:2009} explain, impacts appear because a change of the value of
a covariate in a region will affect not only the region itself (direct impact)
but also other regions indirectly (indirect impact).

For the linear models
presented in Section \ref{sec:spateco}, impacts per covariate $r$ can be defined as

\begin{equation}
\frac{\partial E[y_i \mid x_r]}{\partial x_{jr}}\ \ \ i,j=1,\ldots,n;\ r=1\ldots,p 
\end{equation}
\noindent
with $x_r$ the $r$-th covariate.  This will measure the change in the response
observed in area $i$ when covariate $r$ is changed in area $j$. For the spatial
probit models, the impacts are defined as \citep{LeSageetal:2011}: 

\begin{equation}
\frac{\partial Pr(y_i=1)}{\partial x_{jr}}\ \ \ i,j=1,\ldots,n;\ r=1\ldots,p 
\end{equation}

In both cases, the impacts will produce a $n\times n$ matrix of impacts
$S_r(W)$ for each covariate. The values on the diagonal of this matrix are called {\em direct
impacts}, as they measure the  change in the response when the covariate is
changed at the same area (i.e., the value of covariate $r$ in area $i$ is
changed). In order to give an overall measure of the direct impacts, its 
average is often computed and it is called {\em average direct impact}.

Similarly, {\em indirect impacts} are defined as the off-diagonal elements of
$S_r(W)$, and they measure the change in the response in one area when
changes in covariate $r$ happen at any other area. A global measure of the
indirect impacts is the sum of all off-diagonal elements divided by $n$,
which is called the {\em average indirect impact}.

Finally, the {\em average total impact} is defined as the sum of direct and
indirect impacts. This gives an overall measure of how the response is affected
when changes occur in a covariate at any area.

\citet{Bivandetal:2014} summarise the form of the impacts for different models
and provide some ideas on how to compute the different average impacts with
\pkg{R-INLA} and BMA.  For a Gaussian response, the impacts matrix for the SEM
model is simply a diagonal matrix with coefficient $\beta_r$ in it, i.e.,
$S_r(W)=I_n\beta_r$. For the SDM
model, the impacts matrix is

\begin{equation}
S_r(W)= (I_n-\rholag W)^{-1} (I_n \beta_r+W\gamma_r);\ r=1,\ldots,p.
\label{eq:impactssdm}
\end{equation}
\noindent
The impact matrix for the SLM model is the same as in equation (\ref{eq:impactssdm})
with $\gamma_r=0,\ r=1,\ldots,v$, i.e., the coefficients of the lag covariates
are not considered.

In addition, the impacts matrix for the SDEM model is 

\begin{equation}
S_r(W)= (I_n \beta_r+W\gamma_r);\ r=1,\ldots,p
\label{eq:impactssdem}
\end{equation}
\noindent
Finally, the SLX model shares the same impacts matrix as the SDEM model;
in both cases, the average impacts are the coefficients $\beta_r$
(direct) and $\gamma_r$ (indirect), for which inferences are readily
available.

In the case of the spatial probit, the impacts matrices are similar but they
need to be premultiplied by a diagonal matrix $D(f(\eta))$, which is a $n\times
n$ diagonal matrix with entries $f(\eta_i),i=1,\ldots,n$, where $f(\eta_i)$ is
the standard Gaussian distribution evaluated at the expected value of the
linear predictor of observation $i$.  For example, for the spatial probit SDM
model this is:

\begin{equation}
S_r(W)= D(f(\eta))  (I_n-\rholag W)^{-1} (I_n \beta_r +W\gamma_r);\ r=1,\ldots,v,
\label{eq:impprsdm}
\end{equation}

\noindent
where $\eta$ is defined as

\begin{equation}
\eta = (I_n-\rholag W)^{-1} (X\beta_r +WX\gamma_r);\ r=1,\ldots,v.
\label{eq:eta}
\end{equation}

\noindent
The impacts matrix for other spatial probit models can be derived in a similar
way.

\subsection{Approximation of the impacts}

Average direct, indirect and total impacts can be computed by summing over the
required elements in the impacts matrix (and dividing by $n$).  In a few simple
cases, such as the SEM, SDEM and SLX models, the impacts can be computed with
\pkg{R-INLA} as the impacts are a linear combination of the covariate
coefficients.  In general, the impacts cannot be computed directly with
\pkg{R-INLA} as they are a function on several parameters and INLA only
provides marginal inference. 

From a general point of view, the average impacts can be regarded as the
computation of a functions that involves two or three parameters in the model.
Hence, multivariate posterior inference is required to obtain estimates of the
impacts.  We will try to approximate the posterior marginal of the average
impacts the best we can by sampling from the approximate joint posterior (see
below). 

Let us consider the Gaussian case first. For the SEM, the average direct impact
for covariate $r$ is simply the posterior marginal of coefficient $\beta_r$.
So this is a trivial case and inference is exact. Average indirect impacts are
equal to zero, which makes the average total impact equal to the average direct
impact, i.e., $\beta_r$. 

For the SDM  model, the average total impact is 

\begin{equation}
\frac{1}{1-\rholag}(\beta_r+\gamma_r)
\label{eq:sdmimpact}
\end{equation}
\noindent
Note how this is the product of two terms, one on $\rholag$ and the other
one on $\beta_r+\gamma_r$. In order to estimate the posterior distribution
the average total impact, samples from approximate joint posterior of
the parameters invovled can be drawn from the fitted INLA model
using the \texttt{inla.posterior.sample} function \cite[][Chapter 2]{GomezRubio:2020}.

Regarding the average direct impact for the SDM model, this is

\begin{equation}
n^{-1} tr\Big((I_n-\rholag W)^{-1}\Big)\beta_r +
n^{-1}tr\Big((I_n-\rholag W)^{-1}W\Big)\gamma_r
\end{equation}
\noindent
Again, this expression is a non-linear term that involves several parameters
and its posterior distribution can be approximated by sampling from the
approximate posterior distribution obtained with INLA.

For the SDEM and SLX models, the distribution of the average total impact is the
marginal distribution of $\beta_r+\gamma_r$, whilst the associated  direct
impact is given by 

\begin{equation}
n^{-1} tr\Big(I_n\Big)\beta_r +
n^{-1}tr\Big(W\Big)\gamma_r =\beta_r
\end{equation}
\noindent
Hence, inference on the impacts is exact for the SDEM and SLX models.

In the case of the spatial probit, the average total impacts
are as before but  multiplied by 

\begin{equation}
\sum_{i=1}^n \frac{f(\eta_i)}{n}
\label{eq:Df}
\end{equation}
\noindent
The average direct impact is the trace of $S_r (W)$, which now takes
a more complex form as it involves $D(f(\eta))$, divided
by $n$. For example,
for the SDM model it is 

\begin{eqnarray}
n^{-1} tr[D(f(\eta))  (I_n-\rholag W)^{-1} (\beta_r +W\gamma_r)] =\\\nonumber
n^{-1} tr[D(f(\eta))  (I_n-\rholag W)^{-1}] \beta_r +\\\nonumber 
+n^{-1} tr[D(f(\eta))  (I_n-\rholag W)^{-1}W]\gamma_r)
\label{eq:SDMsppbimp}
\end{eqnarray}
\noindent
The posterior distribution of the impacts will be approximated using sampling from INLA, as stated above.

%
%
%
%

\section{Further topics}

\subsection{Some applications in Spatial Econometrics}
\label{sec:appl}

\citet{LeSagePace:2009} not only describe how to fit Bayesian Spatial
Econometrics models using MCMC but also discuss how to take advantage 
of the Bayesian approach to tackle a number of other issues. In this
section we aim at discussing other applications when dealing
with spatial econometrics models.

\subsubsection{Model selection}
\label{subsec:modsel}

\pkg{R-INLA} reports the marginal likelihood of the fitted model $\mathcal{M}$,
i.e., $\pi(y|\mathcal{M})$, which can be used for model selection, as
described in \citet[][Section 6.3]{LeSagePace:2009} and
\citet{Bivandetal:2014}.  For example, if we have a set of $m$ fitted models
$\{\mathcal{M}_i\}_{i=1}^m$ with marginal likelihoods
$\{\pi(y|\mathcal{M}_i)\}_{i=1}^m$, we may select the model with the highest
marginal likelihood as the ``best'' model. 

Following a fully Bayesian approach, we could compute the posterior probability
of each model taking a set of prior model probabilities
$\{\pi(\mathcal{M}_i)\}_{i=1}^m$ and combining them with the marginal likelihoods using Bayes' rule:

\begin{equation}
\pi(\mathcal{M}_i|y)= \frac{\pi(y|\mathcal{M}_i) \pi(\mathcal{M}_i)}{\sum_{j=1}^m \pi(y|\mathcal{M}_j) \pi(\mathcal{M}_j)}\ \ i=1,\ldots,m
\label{eq:postprobM}
\end{equation}
\noindent
If all models are thought to be equally likely a priori then the priors are
taken as $ \pi(\mathcal{M}_i)= 1/m$, so that the posterior probabilities are
obtained by re-scaling the marginal likelihoods to sum up to one:

\begin{equation}
\pi(\mathcal{M}_i|y)= \frac{\pi(y|\mathcal{M}_i)}{\sum_{j=1}^m \pi(y|\mathcal{M}_j)}\ \ i=1,\ldots,m
\label{eq:postprobM2}
\end{equation}

This model selection approach can be applied to models
with very different structures. They could be models with different
spatial structures,  different latent effects or based on different sets
of covariates. In Section \ref{sec:examples} we show an example
based on comparing models with different spatial structures in the second 
example on the Katrina business data.

In addition, \pkg{R-INLA} implements a number of criteria for model selection,
such as the Deviance Information Criterion
\citep[][DIC]{Spiegelhalteretal:2002} and Conditional Predictive Ordinate
\citep[][CPO]{isi:000291435700007}.  These criteria can be used to compare
different models and perform model selection as well.

\subsubsection{Variable selection}

As a particular application of model selection we will discus here how to deal
with variable selection. In this case, models differ in the covariates that are
included as fixed effects. The number of possible models that appear is usually
very large. For example, 20 covariates will produce $2^{20}$ possible models,
i.e., more than 1 million models to be fitted.  As stated before, posterior
probabilities for each model can be computed using the marginal likelihood as
in equation (\ref{eq:postprobM2}).  In principle, given that \pkg{R-INLA} fits
models very quickly and that a large number of models can be fitted in parallel
on a cluster of computers, it would be feasible to fit all possible models. 

As an alternative approach, stepwise regression can be performed based on any
of the model selection criteria available. In particular, the DIC provides a
feasible way of performing variable selection. This can be included in a
step-wise variable selection procedure which will not explore all possible
models but that can lead to a sub-optimal model.


\subsubsection{Model averaging}

Sometimes, an averaged model may be obtained from other fitted models.  We have
already pointed out how \citet{Bivandetal:2014} use Bayesian model averaging to
fit spatial econometrics models using other models with simpler random effects.
\cite{GomezRubioetal:2020} describe the use of Bayesian model averaging with INLA and how to fit a spatial econometrics models with two spatial autoregression
parameters. This approach can be employed to fit highly parameterized models
with INLA.

However, a BMA approach can also be used to combine different models for other
purposes. For example, when the adjacency matrix is unknown we may fit
different models using slightly different adjacency matrices.
\citet{LeSagePace:2009} discuss BMA in the context of spatial econometrics.  In
Section  \ref{sec:examples} we have considered this in the second example on
the Katrina business data where different spatial structures are considered
using a nearest neighbour algorithm.

\subsection{Other issues in Bayesian inference}
\label{sec:other}

So far, we have made a review of some existing and widely used Spatial
Econometrics models, and how these models can be fitted using INLA and its
associated software \pkg{R-INLA}. Now, we will focus on other general problems
that can be tackled using this new approach.

\subsubsection{Linear combinations and linear constraints on the parameters}

\pkg{R-INLA} allows the computation of posterior marginals of linear
combinations on the latent effects. This can be very useful to compute
some derived quantities from the fitted models, such as some of the
impacts described in Section \ref{sec:impacts}.

Furthermore, \pkg{R-INLA} allows the user to define linear constraints on any
of the latent parameters and other quantities, including the linear predictor.
This is useful, for example, to produce benchmarked estimates, i.e.,
model-based estimates obtained at an aggregation level that must match a
particular value at a different aggregation level.


\subsubsection{Prediction of missing values in the response}
\label{subsec:misval}

Missing values often appear in spatial econometrics because actual data have
not been gathered for some regions or the respondent was not available at the
time of the interview. Sometimes, missing data appear because of the way
surveys are designed as the sample is taken to be representative of the whole
population under study and many small areas may not be sampled at all.
Missing values may also appear in the covariates, but we will only consider
here the case of missing values in the response.

With \pkg{R-INLA}, a posterior marginal distribution will be obtained and
inference and predictions on the missing responses can be made from this. Note
that this is a prediction only and that uncertainty about the missing values
will not influence the parameter estimates.

The case of
missing values in the covariates is more complex.  First of all, we will need
to define a reasonable imputation model and, secondly, the missing values and
the parameters in the new imputation model will be treated as hyperparameters
in our approach, increasing the number of hyperparameters and making a
computational solution  infeasible.

\subsubsection{Choice of the priors}

\citet{LeSagePace:2009}  briefly discuss the choice of different priors for the
parameters in the model, and they stress the importance of having vague priors. 
For the spatial autocorrelation parameter they
propose a uniform distribution in the range of this parameter. A Normal prior
with zero mean and large variance is used for $\beta$. A Gamma prior with small
mean and large variance is proposed for the variance $\sigma^2$. 
However wise these choices may seem, it is not clear how these priors
will impact on the results.

Because of the way different models in the Spatial 
Econometrics Toolbox are implemented, it is difficult to assess the impact of the priors,
as using different priors will require rethinking how the MCMC sampling is
done. New conditional distributions need to be worked out and implemented

\pkg{R-INLA} provides a simple interface with some predefined priors that can
be easily used. Other priors can be defined by using a convenient language
and plugged into the \pkg{R-INLA} software. Hence, it is easier and faster to
assess the impact of different priors.
See Chapter 5 in \cite{GomezRubio:2020} for a general discussion on the use
of priors with \pkg{R-INLA}.

%
%

For example, \citet{Gelman:2006} has suggested that Gamma priors for the
variances were not adequate for the variance parameter in Gaussian models as
they were too informative. Instead, they have proposed the use of a
half-Cauchy distribution. A model with this prior can be easily implemented by
defining the half-Cauchy prior and passing it to \pkg{R-INLA}.

\section{Examples}
\label{sec:examples}

\subsection{Boston housing data}
\label{example:Boston}

\citet{HarrisonRubinfeld:1978} study the median value of owner-occupied
houses in the Boston area using 13 covariates as well. Note that the
median value has been censored at \$50,000 and that we omit tracts that
are censored, leaving 490 observations \citep{PaceGilley:1997}. The
spatial adjacency that we will consider is for census tract contiguities.

\citet{Bivandetal:2014} use INLA and Bayesian model averaging to fit
SEM, SLM and SDM models to this dataset (but including all 506 tracts
and using a different representation of adjacency). Here we will use the
new \code{slm} latent model for \pkg{R-INLA} to fit the same models. In
principle, we should obtain similar results and we will benefit from all
the other built-in features in \pkg{R-INLA} (such as, summary statistics,
model selection criteria, prediction, etc.).

\begin{figure}
\begin{center}
\includegraphics[width=9cm]{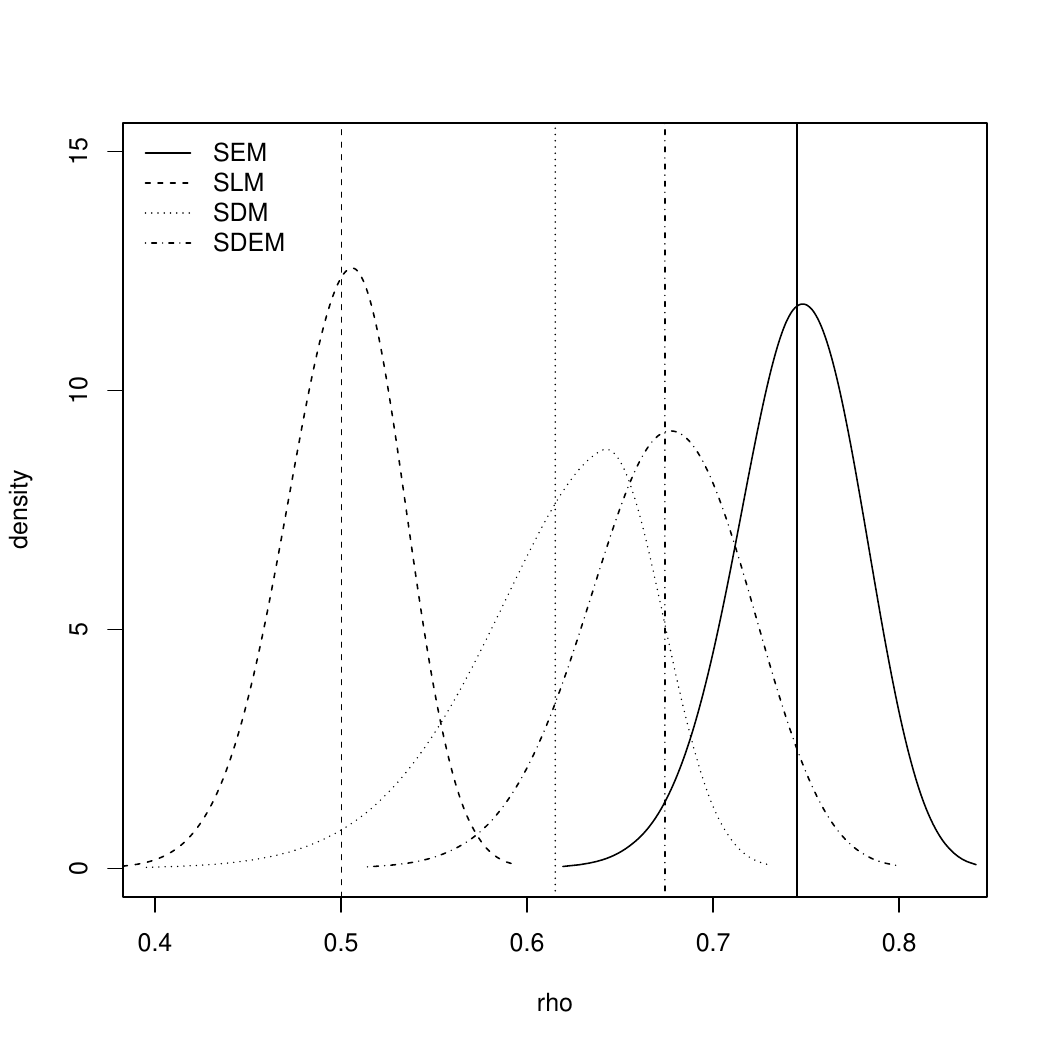}
\end{center}

\caption{Posterior marginal of the spatial autocorrelation parameters
with posterior means (vertical lines),
Boston housing data.}
\label{fig:Brho}
\end{figure}

First of all, we have fitted the five models described in Section 
\ref{sec:spateco}. Point estimates of the fixed effects
are summarised in Table \ref{tab:Bfixed}.
In addition, the posterior marginal of the spatial autocorrelation parameters
have been displayed in Figure \ref{fig:Brho}, including posterior means. Note
that these values are not the ones reported in the \pkg{R-INLA} output and that
we have re-scaled them as explained in Section \ref{sec:slm}. In this case, the 
range used for the spatial autocorrelation parameter is $(-1, 1)$.
This will make the summary statistics for $\rho$ directly comparable to those
reported in \citet{Bivandetal:2014}. 

\begin{figure}
\begin{center}
\includegraphics[width=16cm]{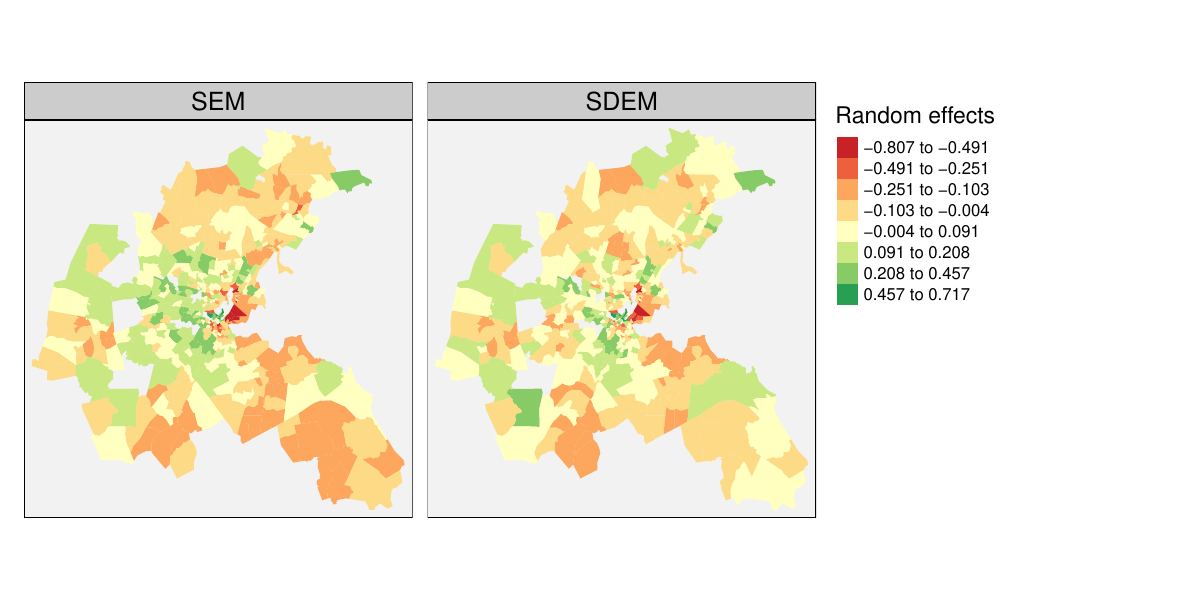}
\end{center}
\caption{Posterior means of the \code{slm} latent effects for the SEM and SDEM
models, Boston housing data.}
\label{fig:slm-effects}
\end{figure}

In general, all our results match theirs
as expected. However, the new \code{slm} latent effects makes fitting these
models with \pkg{R-INLA} simpler.  Finally, Figure~\ref{fig:slm-effects} shows
a map of the values of the \code{slm} latent effects for the SEM and SDEM
models.

Note that in order to fit the model we have set the variance of the
Gaussian likelihood to a fixed and tiny value ($e^{-15}$) because this
error term does not appear in the different spatial econometrics model fit.
A side effect of this is that the DIC will be the same for all models (and it cannot be used for model comparison) and that the fitted values will also have a tiny variance. The fitted values could be computed in the right way by adding extra observations with missing values (i.e., \texttt{NA}) in the response; this obsevations will not be used for model fitting and the fitted values will now account for the required uncertainty.

\begin{table}
\centering

\begin{scriptsize}
\begin{tabular}{rrrrrrrrr}
  \hline
 & SEM & SLM & SDM & SDMlag & SDEM & SDEMlag & SLX & SLXlag \\
  \hline
(Intercept) & 3.542 & 2.168 & 1.946 &  & 4.364 &  & 5.122 &  \\
  CRIM & -0.007 & -0.007 & -0.006 & -0.003 & -0.007 & -0.006 & -0.007 & -0.015 \\
  ZN & 0.000 & 0.000 & 0.000 & -0.000 & 0.000 & -0.001 & 0.000 & 0.000 \\
  INDUS & 0.001 & 0.002 & -0.000 & 0.000 & 0.001 & 0.001 & -0.002 & 0.002 \\
  CHAS1 & -0.047 & -0.002 & -0.062 & 0.110 & -0.047 & 0.133 & -0.065 & 0.194 \\
  I(NOX\verb|^|2) & -0.149 & -0.232 & 0.011 & -0.404 & -0.050 & -0.585 & 0.081 & -1.056 \\
  I(RM\verb|^|2) & 0.010 & 0.008 & 0.010 & -0.009 & 0.010 & -0.002 & 0.009 & -0.007 \\
  AGE & -0.001 & -0.000 & -0.001 & 0.002 & -0.001 & 0.001 & -0.001 & 0.002 \\
  log(DIS) & -0.033 & -0.140 & -0.040 & -0.064 & -0.082 & -0.076 & -0.038 & -0.221 \\
  log(RAD) & 0.059 & 0.062 & 0.052 & -0.004 & 0.059 & 0.014 & 0.043 & 0.078 \\
  TAX & -0.001 & -0.000 & -0.000 & 0.000 & -0.000 & 0.000 & -0.000 & 0.000 \\
  PTRATIO & -0.018 & -0.013 & -0.013 & -0.002 & -0.016 & -0.016 & -0.012 & -0.027 \\
  B & 0.001 & 0.000 & 0.001 & -0.001 & 0.000 & -0.000 & 0.001 & -0.001 \\
  log(LSTAT) & -0.226 & -0.218 & -0.215 & 0.054 & -0.233 & -0.112 & -0.234 & -0.180 \\
   \hline
\end{tabular}
\end{scriptsize}

\caption{Posterior means of the fixed effects coefficients,
Boston housing data.}
\label{tab:Bfixed}
\end{table}

\subsubsection{Smoothing covariates}

\begin{figure}[h!]
\begin{center}
\includegraphics[width=12cm]{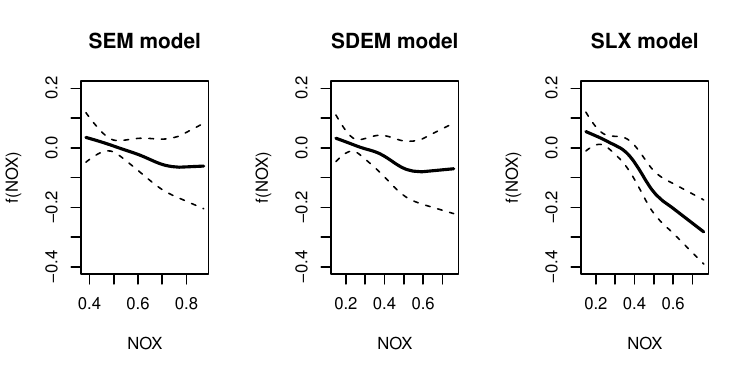}
\end{center}
\caption{Smoothed effects of nitric oxides concentration (NOX) for three
spatial econometrics models with 95\% credible intervals.}
\label{fig:NOX}
\end{figure}

So far we have considered the squared values of NOX in our models but the
relationship between this covariate and the response may take other forms.  In
Section \ref{subsec:addeff} we have discussed how \pkg{R-INLA} implements some
latent effects to smooth covariates \citep[see, for example,][Chapter
9]{GomezRubio:2020}. One of them is the second order random walk that we have
used here to smooth the values of nitric oxides concentration (parts per 10
million), which is covariate NOX.  This smoother needs to be included
additively in all the other effects, so it is only readily available for the
SEM, SDEM and SLX models. This latent model has only an hyperparameter, which
is its precision. In order to avoid overfitting and force the
random walk to produce a smooth function, a Gamma prior with mean 2000 and
variance 10 has been used for the precision in all models, so that the level of
smoothing can be compared.  The results are shown in Figure \ref{fig:NOX}.

SEM and SDEM show a similar linear effect, which is consistent with
the findings in \citet{HarrisonRubinfeld:1978}. The SLX model seems 
to provide similar estimates of this effect but with considerably narrower
credible intervals. As no spatial random
effects are included in this model, we believe that the smoother on NOX
is picking up residual spatial effects.

\subsubsection{Impacts}

Average direct, indirect and total impacts have been computed for the models
fitted to the Boston housing data set. In addition, for the SLM and SDM models
we have also fitted the models using maximum likelihood and computed
their impacts for purposes of comparison. 
Average direct impacts, average total impacts and average indirect impacts are
provided as supplementary materials.  Impacts are very similar between  ML and
Bayesian estimation for the models where we have computed them using 
functioni in the \pkg{spdep} package.

\begin{figure}[h]
\begin{center}
\includegraphics[width=10cm]{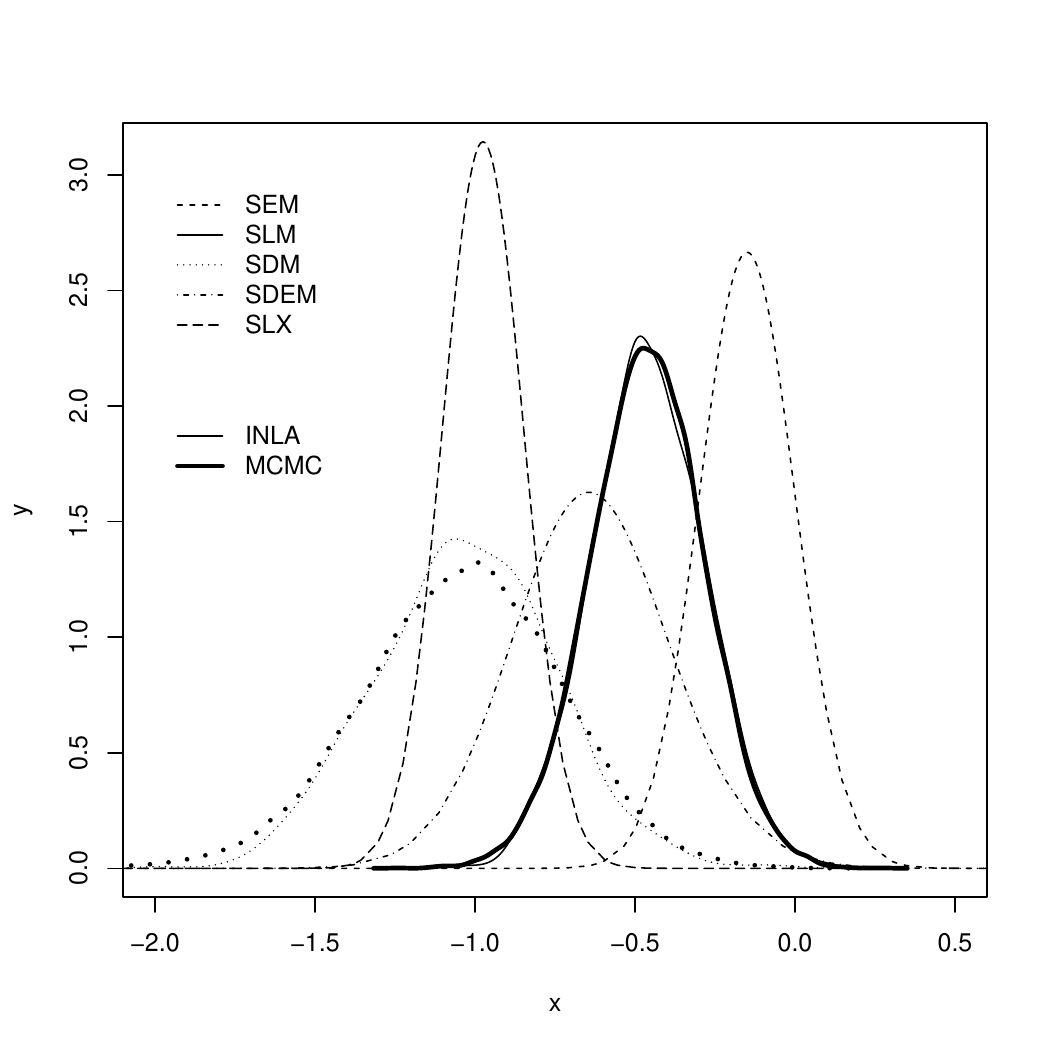}
\end{center}

\caption{Total impacts for NOX-squared obtained with different models, Boston housing data}
\label{fig:NOXimp}
\end{figure}

Inference on the different impacts is based on their respective  posterior
marginal distributions provided by \pkg{R-INLA}. In addition to the posterior
means other statistics can be obtained, such as standard deviation, quantiles
and credible intervals. These posterior marginal distributions can also be
compared to assess how different models produce different impacts.
Figure \ref{fig:NOXimp} shows the total impacts for NOX-squared under
five different models. Given that for models SLM and SDM we were using
an approximation we have included, in a thicker line, the distribution of the total
impacts obtained with the Matlab code in the Spatial Econometrics Toolbox
for these two models.
The results clearly show that our approximation is very close to the 
results based on MCMC. Furthermore, we have checked that similar accuracy
is obtained for all the covariates included in the model.

\subsubsection{Prediction of missing values}

As stated in Section \ref{subsec:misval}, INLA and \pkg{R-INLA} can provide
predictions of missing values in the response. We will use this feature to
provide predictions of the 16 tracts with censored observations of the median
values, treated as missing values. This will allow us to use the
complete adjacency structure and to borrow information from neighbouring areas
to provide better predictions.

\begin{figure}
\begin{center}
\includegraphics[width=14cm]{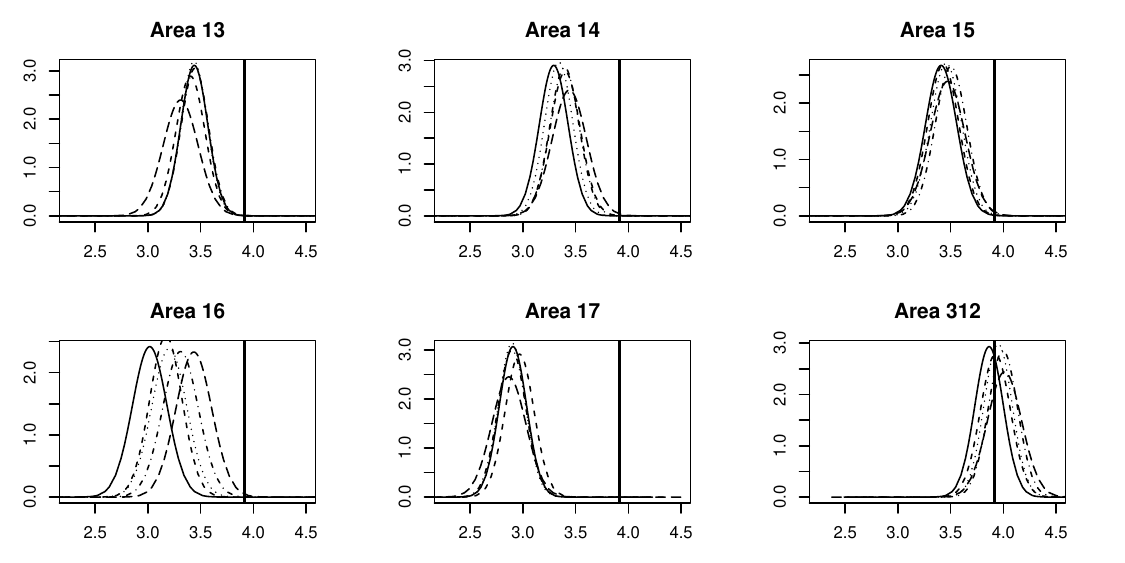}
\end{center}
\caption{Marginal distributions of the median housing values for 6 of the
16 areas with censored observations, Boston housing data with full adjacency matrix. The legend is as in Figure \ref{fig:NOXimp}. Vertical lines shows where the censoring cuts in.}
\label{fig:slm-full-pred}
\end{figure}

With \pkg{R-INLA}, this is as simple as setting the censored values
in the response to \code{NA}. We will obtain a predictive distribution
for the missing values so that inference can be made from them. We have
represented the five marginal distributions obtained with the models in
Figure~\ref{fig:slm-full-pred} for 6 selected areas. The vertical line shows where the censoring cuts in.

Areas 13 to 17 seem to have predicted values well below the cut-off point.
These ares are located in the city center, where house prices are likely to be
higher than average and that is why our model does not predict well there.
Furthermore, predicted values the remaining 11 areas with missing values have a
similar behaviour as in Area 312 (included in Figure~\ref{fig:slm-full-pred}),
that is, the cut-off point is close to the median of the predicted values.



Furthermore, in the supplementary materials we show the posterior means  of the
\code{slm} latent effects for the SEM and SDEM models 
in the same way as in Figure \ref{fig:slm-effects} for the incomplete data set.
The main difference is that the new maps include predictions for the areas with
the censored observations but the estimates in the common tracts are similar.

\subsection{Business opening after Katrina}
\label{example:Katrina}

\citet{LeSageetal:2011} study the probabilities of reopening a business in New
Orleans in the aftermath of hurricane Katrina. They have used a spatial probit,
as the one described in equation (\ref{eq:probit}). Here we reproduce the
analysis with a continuous link function (i.e., a probit function) and the new
\code{slm} latent model, similarly as in \citet{Bivandetal:2014}.

\subsubsection{Standard models}

Table~\ref{tab:Kfixed} shows point estimates (posterior means) of the fixed
effects for the five of the models discussed in Section \ref{sec:spateco}.

\begin{figure}[h!]
\begin{center}
\includegraphics[width=9cm]{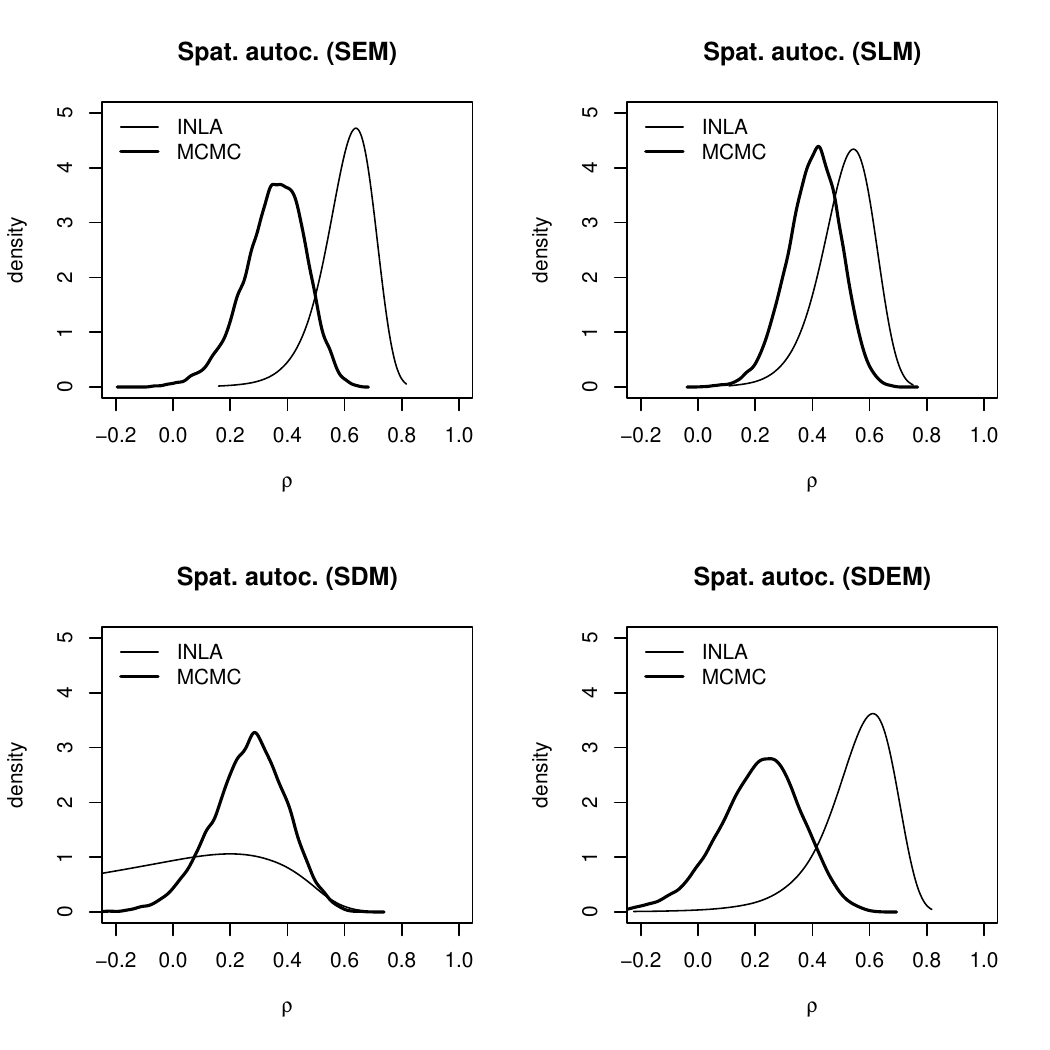}
\end{center}
\caption{Posterior marginal of the spatial autocorrelation parameters
for different models for INLA and MCMC, Katrina business data. Differences in the estimates can be explained by the different link functions used in the models fit with MCMC and INLA.}
\label{fig:Krho}
\end{figure}

Furthermore, Figure~\ref{fig:Krho} shows the marginal distribution of the
spatial autocorrelation parameters of four different spatial econometrics
models for INLA and MCMC. For the INLA models, the values of the spatial autocorrelation
parameters have been properly re-scaled to fit in the correct range and not
constrained to the (0,1) interval. In this example $\rho$ could be betwee from $-3.276$
to $1$, but we have only considered it to be in the $(-1, 1)$ so that a fair
comparisson with MCMC can be done. INLA estimates for both fixed and spatial autocorrelation estimates are very similar to
those reported in \citet{Bivandetal:2014}. The posterior mean of the spatial
autocorrelation for the SDM differs but this may be because
\citet{Bivandetal:2014} constrain $\rho$ to be in the (0,1) interval.  The SDM
model seems to have a weaker residual spatial correlation, probably because the
inclusion of the lagged covariates reduces the autocorrelation in the response.

Regarding INLA and MCMC estimates, although the marginals are close for the SLM model, they are a bit further away for all the other three
models presented in Figure ~\ref{fig:Krho}. The posterior modes are close but
these differences may be due to the fact that two different link functions are used as the MCMC implementation defines a latent continuous variable $y_i^*$
so that the response is 1 when $y^*_i$ is non-negative and 0 otherwise. This makes the models fit with INLA and MCMC different in practice.

%
%


\begin{table}[ht]
\centering
\begin{scriptsize}
\begin{tabular}{rrrrrrrrr}
  \hline
 & SEM & SLM & SDM & SDMlag & SDEM & SDEMlag & SLX & SLXlag \\
  \hline
(Intercept) & -19.800 & -8.142 & -15.108 &  & -12.775 &  & -9.116 &  \\
  flood\_depth & -0.434 & -0.187 & -0.480 & 0.012 & -0.497 & 0.100 & -0.323 & 0.048 \\
  log\_medinc & 1.925 & 0.776 & 2.329 & -0.891 & 2.092 & -0.887 & 1.453 & -0.582 \\
  small\_size & -0.348 & -0.403 & -0.431 & -1.113 & -0.403 & -0.583 & -0.287 & -0.703 \\
  large\_size & -0.403 & -0.475 & -0.372 & -1.646 & -0.416 & -1.913 & -0.242 & -1.011 \\
  low\_status\_customers & -0.335 & -0.480 & -0.103 & -1.759 & -0.089 & -1.620 & -0.059 & -1.083 \\
  high\_status\_customers & 0.135 & 0.105 & 0.049 & 0.113 & 0.067 & 0.261 & 0.035 & 0.083 \\
  owntype\_sole\_proprietor & 0.774 & 0.790 & 0.910 & 1.393 & 0.844 & 0.853 & 0.613 & 0.808 \\
  owntype\_national\_chain & 0.087 & 0.138 & 0.213 & 2.066 & 0.174 & 2.449 & 0.119 & 1.167 \\
   \hline
DIC & 664.563 & 664.358 & \multicolumn{2}{c}{683.732} & \multicolumn{2}{c}{664.680} &  \multicolumn{2}{c}{703.411}\\
M. Lik. & -386.500 & -401.648 & \multicolumn{2}{c}{-435.459} & \multicolumn{2}{c}{-408.367} & \multicolumn{2}{c}{-410.769}  \\
   \hline
\end{tabular}
\end{scriptsize}
\caption{Summary of point estimates of the fixed effects coefficients, Katrina business data.}
\label{tab:Kfixed}
\end{table}

\subsubsection{Exploring the number of neighbours}

\citet{LeSageetal:2011} use a nearest neighbours method to obtain an adjacency
matrix for the businesses in the dataset. Also, they have explored the optimal
number of nearest neighbours by fitting the model using different numbers of
nearest neighbours and using the model with the lowest DIC
\citep{Spiegelhalteretal:2002} as the one with the optimal number of
neighbours. For the 3-month horizon model, they compared a window of 8-14
neighbours, probably because of the computational burden of MCMC, which they
used to fit their models.

\begin{figure}
\begin{center}
\includegraphics[scale=.4]{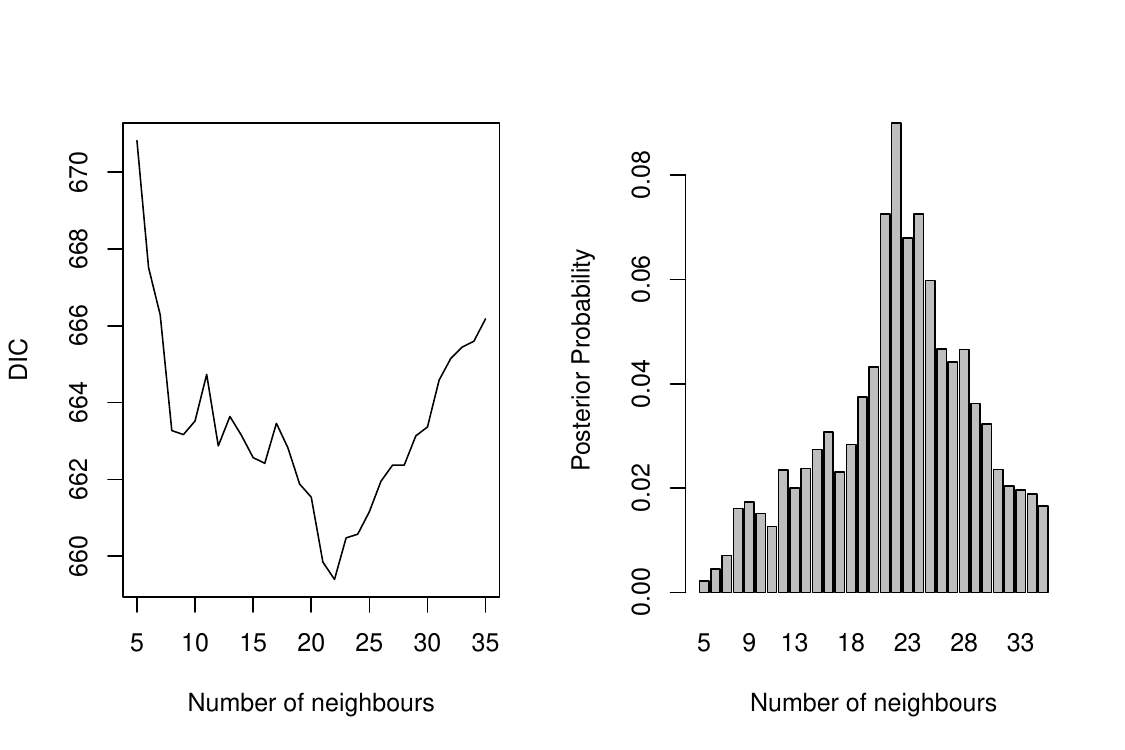}
\includegraphics[scale=.4]{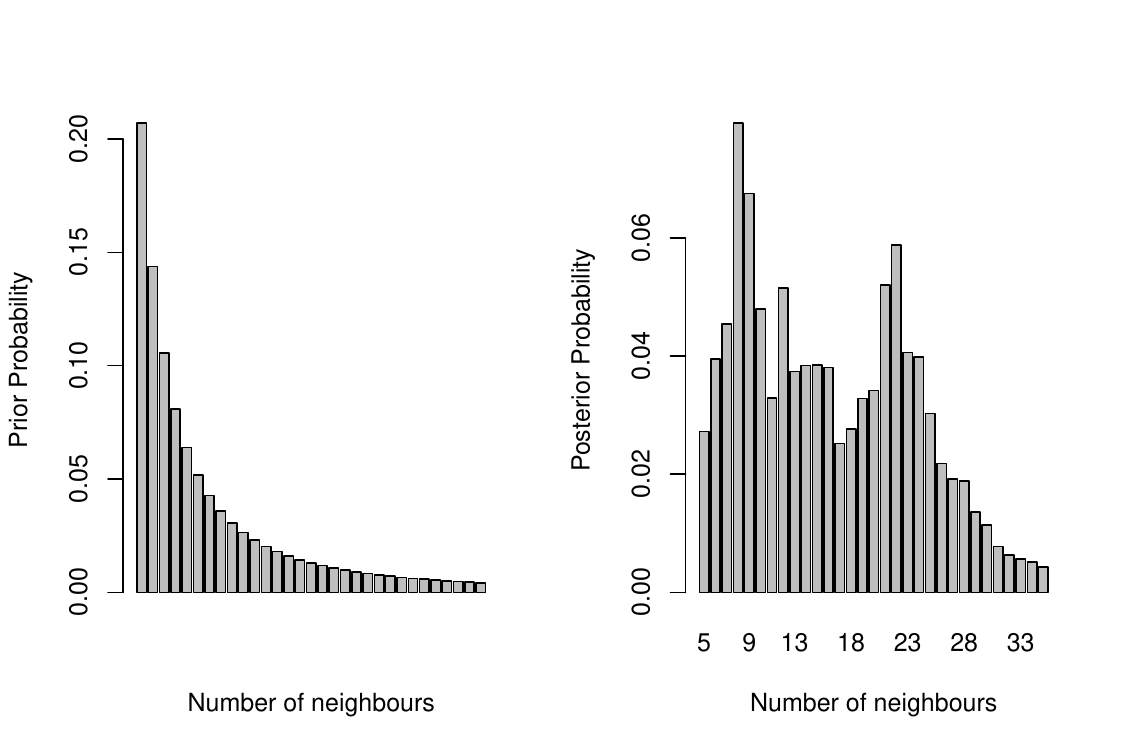}
\end{center}

\caption{DIC (top-left) and posterior probabilities (top-right)  for spatial probit
models with different adjacency structures based on a nearest neighbours
approach using a uniform prior, Katrina business data.
Prior (bottom-left) and posterior (bottom-right) probabilities for spatial probit
models with different adjacency structures based on a nearest neighbours
approach using an informative prior, Katrina business data.}

\label{fig:katneigh}
\label{fig:katneighprior}
\end{figure}

The newly available \code{slm} model in \pkg{R-INLA} makes exploring the
optimal number of neighbours faster and simpler. We have increased
the number of neighbours considered, between 5 and 35, and fitted the SLM
model using different adjacency matrices. These adjacency matrices
have been created using the nearest neighbour algorithm with different
values of the number of neighbours. Figure \ref{fig:katneigh} shows how
the optimal number of neighbours seems to be 22 according to both the
DIC and the posterior probability (as explained in Section \ref{sec:appl})
criteria. However, we believe that this should be used as a guidance to set the
number of neighbours as there may be other factors to take into account. In
particular, a nearest neighbour approach may consider as neighbours businesses that are in different parts of the city, particularly if the number of nearest
neighbours is allowed to be high.

When computing the posterior probability, the prior probability of each model is
taken so that $\pi(\mathcal{M}_i)\propto 1$, and  there is no prior preference
on the number of neighbours.  If we decide to favour adjancencies based on a
small number of neighbours we may use a more informative prior. For example, we
could take $\pi(k)\propto \frac{1}{k^2}$ so that neighbourhoods with a smaller
number of neighbours are preferred. Figure \ref{fig:katneighprior} shows the
prior and posterior probability  for different values of the number of
neighbours. Now it can be seen how our prior information produces  different
posterior probabilities, with an optimal number of neighbours of 8.

%

Finally, it is also possible to average over the ensemble of models using
Bayesian model averaging. This will produce a fitted model that takes into
account all the adjacency structures, weighted according to their marginal
likelihoods. Table \ref{tab:BMAneigh} shows the estimates of the fixed effects
for the model with the highest posterior probability according to a uniform
prior ($\pi(k)\propto 1$) and an informative prior ($\pi(k)\propto 1/k^2$), and
the estimates obtained by averaging over all the fitted models. These models should take into account the uncertainty about the
number of neighbours and can provide different estimates of the fixed effects.
As it happens, the posterior means and standard deviations are slightly 
different, but we can observe higher differences in the case of an informative
prior.

\begin{table}[ht]
\centering
\begin{scriptsize}
\begin{tabular}{r|rr|rr|rr|rr}
\hline
 & \multicolumn{4}{c|}{Model with highest post. prob} & 
 \multicolumn{4}{c}{BMA models} \\
\hline
 & \multicolumn{2}{c}{Uniform prior} & \multicolumn{2}{c|}{Informative prior} & 
 \multicolumn{2}{c}{Uniform prior} & \multicolumn{2}{c}{Informative prior}  \\
  \hline
 & mean & sd & mean & sd & mean & sd & mean & sd \\ 
  \hline
(Intercept) & -6.831 & 2.880 & -8.645 & 3.086 & -7.452 & 3.109 & -8.544 & 3.299 \\
  flood\_depth & -0.130 & 0.054 & -0.206 & 0.056 & -0.155 & 0.067 & -0.201 & 0.070 \\
  log\_medinc & 0.643 & 0.281 & 0.829 & 0.301 & 0.708 & 0.304 & 0.819 & 0.323 \\ 
  small\_size & -0.409 & 0.187 & -0.407 & 0.186 & -0.421 & 0.188 & -0.410 & 0.187 \\
  large\_size & -0.419 & 0.441 & -0.545 & 0.449 & -0.457 & 0.445 & -0.509 & 0.449 \\
  low\_status\_customers & -0.436 & 0.207 & -0.492 & 0.208 & -0.450 & 0.212 & -0.497 & 0.213 \\
  high\_status\_customers & 0.099 & 0.171 & 0.124 & 0.171 & 0.108 & 0.172 & 0.107 & 0.172 \\
  owntype\_sole\_proprietor & 0.805 & 0.262 & 0.757 & 0.262 & 0.788 & 0.262 & 0.772 & 0.262 \\
  owntype\_national\_chain & 0.085 & 0.503 & 0.107 & 0.496 & 0.106 & 0.500 & 0.116 & 0.496 \\
   \hline
\end{tabular}
\end{scriptsize}
\caption{Summary statistics of the covariate coefficients for the model
with the highest probability (under two different priors) and the BMA
model (under two different priors.)}
\label{tab:BMAneigh}
\end{table}

\subsubsection{Impacts}

We have followed the method described in Section \ref{sec:impacts} to
approximate the impacts for the models fitted to the Katrina dataset.  In this
case, we do not have any model fitted using ML with which to compare.  The
implementation of the spatial probit in the Spatial Econometrics Toolbox is for
a different link, so our results cannot directly be compared to MCMC
as reported in \citet{LeSageetal:2011}.  
Direct impacts are shown in Table~\ref{tab:dikatrina}, whilst
total and indirect impacts are available as supplementary materials.

\begin{table}[ht]
\centering
\begin{scriptsize}
\begin{tabular}{rrrrrr}
  \hline
 & INLASEM & INLASLM & INLASDM & INLASDEM & INLASLX \\ 
  \hline
flood\_depth & -0.086 & -0.038 & -0.100 & -0.099 & -0.090 \\ 
  log\_medinc & 0.381 & 0.159 & 0.504 & 0.409 & 0.386 \\ 
  small\_size & -0.070 & -0.084 & -0.097 & -0.079 & -0.079 \\ 
  large\_size & -0.078 & -0.100 & -0.085 & -0.080 & -0.068 \\ 
  low\_status\_customers & -0.067 & -0.100 & -0.034 & -0.018 & -0.016 \\ 
  high\_status\_customers & 0.027 & 0.022 & 0.009 & 0.014 & 0.010 \\ 
  owntype\_sole\_proprietor & 0.154 & 0.165 & 0.201 & 0.165 & 0.169 \\ 
  owntype\_national\_chain & 0.016 & 0.025 & 0.060 & 0.033 & 0.033 \\ 
   \hline
\end{tabular}
\end{scriptsize}
\caption{Direct impacts, Katrina data set.} 
\label{tab:dikatrina}
\end{table}

\begin{figure}[h!]
\includegraphics[width=12cm]{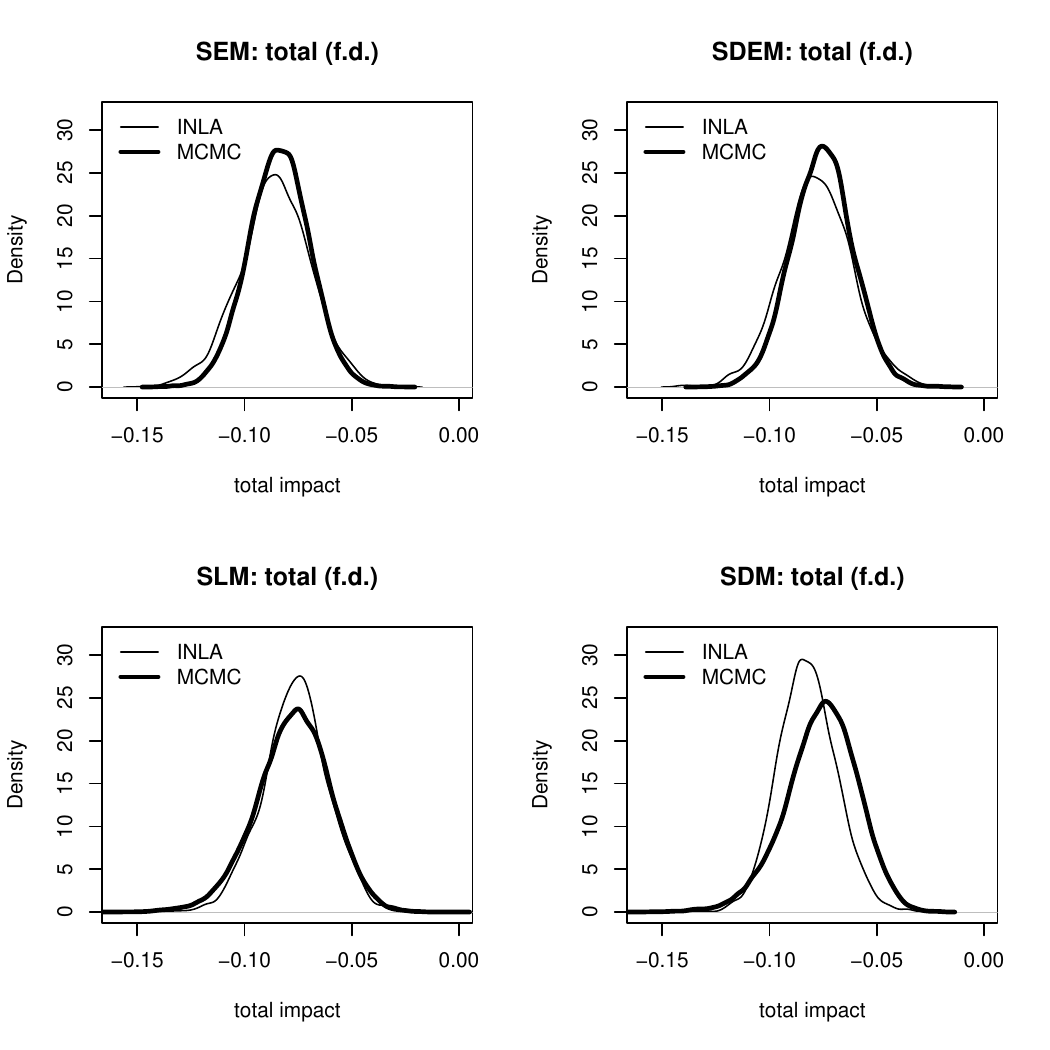}
\caption{Average total impacts for flood depth for the Katrina dataset.}
\label{fig:fdimp}
\end{figure}

Inference on the impacts relies on sampling from the approximation to the joint
posterior distribution which is then used to compute the impacts and estimate
their posterior distributions, as we have already seen in the Boston housing
data example. Figure~\ref{fig:fdimp} shows the estimates of the average total
impacts for flood depth for four models. In a thicker line we have included the
posterior marginal of the impacts computed using the output from MCMC using the
\pkg{spatialprobit} package. As previously stated, MCMC results  can be roughly
compared to our SLM model estimates but keeping in mind that different link
functions have been used and that differences may appear. In this case, the
quality of our approximations differ with the models.  We have found similar
accuracy for all the other variables, which means that our approximation
appears to be acceptable.

Note that the impacts estimated with INLA and MCMC are close
regardless of the estimates of the spatial autocorrelation parameters shown in
Figure~\ref{fig:Krho}.  We believe that this is because the impacts themselves
build on the fitted model parameter values rather than the posterior
distributions, which only enter into simulations to get to the distributions of
the impacts.

\section{Discussion and final remarks}
\label{sec:disc}

In this paper we have described how the analysis of spatial econometrics data
requires the use of very specific models  and how the integrated nested Laplace
approximation  offers an alternative to model fitting. Instead of resorting to
MCMC methods, INLA aims at providing approximate inference on the marginal
distribution of the model parameters. This methodology is implemented in the
\RR package \pkg{R-INLA}, which includes a particular latent effect called
\code{slm} which can be used to fit many spatial econometrics models.

We have also shown how impacts can be approximated when models are fitted with
\pkg{R-INLA}. As this requires multivariate posterior inference, the estimation
of the impacts is achieved by sampling from the approximation to the joint
posterior of the required coefficients and spatial autocorrelation parameter
and computing the impacts using these samples. It is worth noting that these samples are independent, so less samples are required for inference than with
typical MCMC algorithms.

It should be noted that are several the advantages of using INLA and
\pkg{R-INLA}. Not only model specification and fitting is very easy using \RR
but also the computational speed allows us to explore a large number of models.
When the model is not available within the range of latent models available in
\pkg{R-INLA} it is often possible to fit conditional models, on one or two
model parameters,  and then obtained the desired model by averaging over these
models. Furthermore, other important topics in Bayesian inference, such as
prediction of missing responses, model selection and variable selection can be tackled with INLA.

In the future, we expect to explore how to increase the number of Spatial
Econometrics models available in \pkg{R-INLA} and how to extend them. In
particular, we find that there is interesting work to do on models with more
than one spatial term, spatio-temporal models,  the analysis of panel data and
how to account for measurement error in covariates, for example.

\section{Acknowledgments}

V. G\'omez-Rubio has been supported by grants SBPLY/17/180501/000491
(Consejer\'ia de Educaci\'on, Cultura y Deportes, JCCM, Spain, and FEDER) and
PID2019‐106341GB‐I00 (Ministerio de Ciencia e Innovación, Spain).


\bibliographystyle{apalike}
\bibliography{INLA_econometrics,RS_handbook_SpatStat_INLA}

\newpage 

\appendix

\section{\RR interface}
\label{app:Rinterface}

The implementation of the \code{slm} latent model in \pkg{R-INLA} requires
careful attention to the different parameters included in the model. 
It can be defined using the \code{f()} function in \pkg{R-INLA} as

\code{f(idx, model="slm", args.slm=list(rho.min , rho.max, W, X, Q.beta),
   hyper)}

\code{idx} is an index to identify the region, and it can take any values,
\code{model="slm"} indicates that we are using a \code{slm} latent effect,
\code{args.slm} sets some of the data required to fit the model (including the
prior on $\beta$) and \code{hyper} is used to set the prior distributions
for $\tau$ and $\rho$.

\code{rho.min} and \code{rho.max} indicate the range of the spatial
autocorrelation parameter $\rho$. This range will depend on the spatial
weights matrix defined in \code{W}. See for example \citet{Haining:2003} for details.

\code{W} is an adjacency matrix that defines the spatial structure of the data.
In spatial econometrics \code{W} is often taken to be row-standardised.
\pkg{R-INLA} can handle sparse matrices, as defined in the \pkg{Matrix}
package, to make model fitting faster.

\code{X} is the matrix of column-wise covariates. If an intercept is required
in the model, it must be included as a column of 1's (possibly, in the first
column). \code{Q.beta} is a precision matrix for the vector of coefficients
$\beta$. In principle, it can take any form but we advise taking diagonal
matrix with very small values in the diagonal.

In order to set $\beta=0$ we have considered a model with a matrix with zero
columns in \RR. This will mimic a model with no covariates at all.

If the covariates have very different scales it may important to re-scale
them. Otherwise, \pkg{R-INLA} may find some computational problems that may
prevent it from fitting the model. This is particularly important for large
datasets.

\code{hyper} can be used to assign prior distributions to $\tau$
and $\rho$ in the same way as the \code{hyper} parameter in the \code{f()}
functions and that is described in the \pkg{R-INLA} documentation.

The posterior marginal of $\rho$ is reported (and constrained) on the interval
(0,1) and not in the range defined by \code{rho.min} and \code{rho.max}. Hence,
in order to make an appropriate interpretation of the results it must be
linearly transformed to fit in the  (\code{rho.min}, \code{rho.max}) interval
using, for example, \code{inla.tmarginal()}.  Note that if an initial value is
assigned to $\rho$ this must be re-scaled to be in the (0,1) interval as well.

We will not include here any example with \RR code on the use of this new
latent class. The \RR files used to calculate the examples in Section 
\ref{sec:examples} are distributed as supplementary materials to this paper.

\newpage

\section{Expression of \code{slm} as a Gaussian Markov Random Field}
\label{app:GMRF}

In this Appendix we will show how the newly defined \code{slm} latent
effect can be expressed as a Gaussian Markov Random Field. We will
denote the vector of random effects as

$$
\mathbf{x} = (I_n-\rho W)^{-1} (X\beta+\varepsilon)
$$
\noindent
Here $\beta$ has a Gaussian prior with zero mean and precision matrix $Q$, and
$\varepsilon$ a Gaussian distribution with zero mean and precision matrix $\tau
I_n$, with $\tau$ a precision parameter. $Q$ is fixed when the latent effects
are defined, so we will treat it as constant.  Although not explicitly written
down, we are also conditioning on hyperparameters $\tau$ and $\rho$ in all the
distributions that appear below.


Internally, \pkg{R-INLA} works with the joint distribution of $\mathbf{x}$
and $\beta$, denoted by $[\mathbf{x}, \beta]$. We will show
here that this can be expressed as a GMRF with a  sparse precision matrix,
so that it conforms with the INLA framework.

First of all, we will work out the conditional distribution of $\mathbf{x}$
on $\beta$, denoted by $[\mathbf{x} | \beta]$. We will use this later
because $[\mathbf{x}, \beta] = [\mathbf{x} | \beta][\beta]$.

We are assuming that the
joint distribution is Gaussian and, hence, the conditional distribution 
$[\mathbf{x} | \beta]$ is
also Gaussian, with

$$
M = E[\mathbf{x} | \beta] = (I_n-\rho W)^{-1} X\beta
$$
\noindent
and

\begin{eqnarray}
Var[\mathbf{x} | \beta] = Var[(I_n-\rho W)^{-1}X\beta+ (I_n-\rho W)^{-1}\varepsilon|\beta] = \\ \nonumber
(I_n-\rho W)^{-1} Var[\varepsilon|\beta]  ((I_n-\rho W)^{-1}  )' = \\ \nonumber
(I_n-\rho W)^{-1} \frac{1}{\tau} I_n  ((I_n-\rho W)^{-1}  )' = \\ \nonumber
\frac{1}{\tau} (I_n-\rho W)^{-1} (I_n-\rho W')^{-1}   
\end{eqnarray}

The conditional precision can be expressed as

$$
T = Prec[\mathbf{x} | \beta] = \tau (I_n-\rho W')(I_n-\rho W)
$$
\noindent
which is symmetric and highly sparse. 

Hence, we can derive the joint distribution of $\mathbf{x}$ and $\beta$ as

\begin{eqnarray}
\label{eq:jointxb}
[\mathbf{x}, \beta] = [\mathbf{x} | \beta] [\beta] \propto
\exp\{-\frac{1}{2} (\mathbf{x}-M)' T (\mathbf{x}-M)\} 
\exp\{-\frac{1}{2}\beta ' Q \beta\}=\\ \nonumber
\exp\{-\frac{1}{2}\Big(
\mathbf{x}' T\mathbf{x}- \mathbf{x}' T M- M' T\mathbf{x}+ M' T M+ \beta ' Q \beta
\Big)\} = \\ \nonumber
\exp\{-\frac{1}{2} (\mathbf{x}, \beta)' P (\mathbf{x}, \beta)\}
\end{eqnarray}
\noindent
Here $P$ is the precision matrix of $[\mathbf{x}, \beta]$, which is given
by

\begin{eqnarray}
P=
\left[
\begin{array}{cc}
T  & -T (I_n-\rho W)^{-1} X\\
-X' (I_n-\rho W')^{-1} T & Q +\tau X'X\\
\end{array}
\right]
=\\ \nonumber
\left[
\begin{array}{cc}
\tau (I_n-\rho W')(I_n-\rho W)   & -\tau (I_n-\rho W') X\\
-\tau X' (I_n-\rho W) & Q +\tau X'X\\
\end{array}
\right]
\end{eqnarray}

Note that to obtain the previous result we have used  that

\begin{eqnarray}
\mathbf{x}' T M = \mathbf{x}' T (I_n-\rho W)^{-1} X\beta = \\ \nonumber
\mathbf{x}' ( \tau (I_n-\rho W')(I_n-\rho W) ) (I_n-\rho W)^{-1} X\beta = \\ \nonumber
\tau \mathbf{x}' (I_n-\rho W') X \beta
,
\end{eqnarray}
\noindent

$$
M' T \mathbf{x} = (\mathbf{x}' T M)' = \tau \beta' X' (I_n-\rho W) \mathbf{x},
$$
\noindent
and

\begin{eqnarray}
M' T M = \beta' X' (I_n-\rho W')^{-1} T (I_n-\rho W)^{-1} X\beta =\\ \nonumber
 \beta' X'(I_n-\rho W')^{-1}(  \tau (I_n-\rho W')(I_n-\rho W)  ) (I_n-\rho W)^{-1} X\beta = \\ \nonumber
\tau \beta' X'X\beta 
\end{eqnarray}

Furthermore, from the final expression in equation (\ref{eq:jointxb}) it is 
easy to see that the expectation of $[\mathbf{x}, \beta]$
is 

$$
E[\mathbf{x}, \beta] = 0
$$

Hence, the expression of  
$[\mathbf{x}, \beta]$ 
as a Gaussian Markov Random Field
has zero mean and precision matrix $P$. Note how $P$ is a block-matrix which
involves very sparse matrices, which allows for the use of the 
efficient algorithms
described in \citet{RueHeld:2005} for fast computation on GMRF. 

Although $Q$ can take any form, assuming that the coefficients are independent
a priori will lead to a diagonal matrix, which is also sparse. Finally, it may
be worth rescaling the covariates to avoid numerical problems. Including lagged
covariates may lead to further numerical instability as they may be highly
correlated with the original covariates.

\newpage

\section*{Supplementary materials}
\label{app:supmat}

We have prepared a number of \RR files to be distributed with this paper.
These files show how to obtain the results that we have presented here.
Note that all these examples require the use of several \RR packages.
Data are available from different \RR packages and the scripts to reproduce the
results in the paper are at \url{https://github.com/becarioprecario/slm}.
We provide here a list of these files and a short description:


\begin{itemize}

\item \texttt{boston-slm.R}

Analysis of the Boston housing data set using the main spatial econometrics
models.

%



%

\item \texttt{boston-slm-impacts.R}

Computation of the impacts for the Boston housing data example.

\item \texttt{boston-slm-full.R}

Analysis of the Boston housing data set using the main spatial econometrics
models and the full adjacency matrix to perform prediction on the missing
values.

\item \texttt{katrina-slm.R}

Analysis of the Katrina business data using the main spatial econometrics
models with a spatial probit.

\item \texttt{katrina-slm-neigh.R}

Selection of the number of optimal nearest neighbours for the adjacency 
matrix using the Katrina business data.

\item \texttt{katrina-slm-impacts.R}

Computation of the impacts for the Katrina business data example.

\end{itemize}

\newpage

\section*{Boston housing data: supplementary materials}

\begin{table}[h!]

\centering
\begin{tabular}{rrrrr}
  \hline
 & SEM & SLM & SDM & SDEM \\
  \hline
mean & 0.745 & 0.497 & 0.613 & 0.674 \\
  sd & 0.034 & 0.031 & 0.042 & 0.043 \\
  0.025 quantile& 0.675 & 0.432 & 0.531 & 0.585 \\
  0.975 quantile & 0.807 & 0.555 & 0.696 & 0.754 \\
   \hline
\end{tabular}

\caption{Summary of the spatial autocorrelation parameters,
Boston housing data.}
\label{tab:Brho}
\end{table}

\begin{table}[ht!]
\centering
\begin{scriptsize}
\begin{tabular}{rrrrrrr}
  \hline
 & MLSLM & INLASLM & MLSDM & INLASDM & INLASDEM & INLASLX \\ 
  \hline
CRIM & -0.008 & -0.008 & -0.007 & -0.007 & -0.007 & -0.007 \\ 
  ZN & 0.000 & 0.000 & 0.000 & 0.000 & 0.000 & 0.000 \\ 
  INDUS & 0.002 & 0.002 & 0.000 & 0.000 & 0.001 & -0.002 \\ 
  CHAS1 & -0.003 & -0.002 & -0.049 & -0.050 & -0.047 & -0.065 \\ 
  I(NOX\verb|^|2) & -0.246 & -0.249 & -0.061 & -0.060 & -0.050 & 0.081 \\ 
  I(RM\verb|^|2) & 0.008 & 0.008 & 0.010 & 0.010 & 0.010 & 0.009 \\ 
  AGE & -0.000 & -0.000 & -0.001 & -0.001 & -0.001 & -0.001 \\ 
  log(DIS) & -0.149 & -0.149 & -0.056 & -0.050 & -0.082 & -0.038 \\ 
  log(RAD) & 0.066 & 0.066 & 0.057 & 0.057 & 0.059 & 0.043 \\ 
  TAX & -0.000 & -0.000 & -0.000 & -0.000 & -0.000 & -0.000 \\ 
  PTRATIO & -0.013 & -0.013 & -0.015 & -0.015 & -0.016 & -0.012 \\ 
  B & 0.000 & 0.000 & 0.001 & 0.001 & 0.000 & 0.001 \\ 
  log(LSTAT) & -0.231 & -0.231 & -0.229 & -0.230 & -0.233 & -0.234 \\ 
   \hline
\end{tabular}
\end{scriptsize}

\caption{Direct impacts, Boston housing data.}
\label{tab:diboston}
\end{table}

\begin{table}[ht!]
\centering
\begin{scriptsize}
\begin{tabular}{rrrrrrr}
  \hline
 & MLSLM & INLASLM & MLSDM & INLASDM & INLASDEM & INLASLX \\ 
  \hline
CRIM & -0.014 & -0.014 & -0.024 & -0.024 & -0.013 & -0.022 \\ 
  ZN & 0.001 & 0.001 & 0.001 & 0.001 & -0.000 & 0.001 \\ 
  INDUS & 0.004 & 0.004 & 0.001 & 0.000 & 0.001 & 0.000 \\ 
  CHAS1 & -0.005 & -0.005 & 0.127 & 0.127 & 0.086 & 0.129 \\ 
  I(NOX\verb|^|2) & -0.461 & -0.467 & -1.033 & -1.029 & -0.634 & -0.975 \\ 
  I(RM\verb|^|2) & 0.015 & 0.015 & 0.004 & 0.003 & 0.009 & 0.003 \\ 
  AGE & -0.000 & -0.000 & 0.001 & 0.001 & -0.000 & 0.001 \\ 
  log(DIS) & -0.279 & -0.282 & -0.273 & -0.274 & -0.158 & -0.259 \\ 
  log(RAD) & 0.123 & 0.124 & 0.125 & 0.126 & 0.073 & 0.121 \\ 
  TAX & -0.001 & -0.001 & -0.000 & -0.000 & -0.000 & -0.000 \\ 
  PTRATIO & -0.025 & -0.025 & -0.039 & -0.039 & -0.031 & -0.038 \\ 
  B & 0.000 & 0.000 & 0.000 & 0.000 & 0.000 & -0.000 \\ 
  log(LSTAT) & -0.434 & -0.434 & -0.424 & -0.425 & -0.345 & -0.414 \\ 
   \hline
\end{tabular}
\end{scriptsize}

\caption{Total impacts, Boston housing data.}
\label{tab:tiboston}
\end{table}

\begin{table}[ht!]
\centering
\begin{scriptsize}
\begin{tabular}{rrrrrrr}
  \hline
 & MLSLM & INLASLM & MLSDM & INLASDM & INLASDEM & INLASLX \\ 
  \hline
CRIM & -0.007 & -0.007 & -0.016 & -0.016 & -0.006 & -0.015 \\ 
  ZN & 0.000 & 0.000 & 0.000 & 0.000 & -0.001 & 0.000 \\ 
  INDUS & 0.002 & 0.002 & 0.001 & 0.000 & 0.001 & 0.002 \\ 
  CHAS1 & -0.002 & -0.002 & 0.176 & 0.177 & 0.133 & 0.194 \\ 
  I(NOX\verb|^|2) & -0.216 & -0.218 & -0.973 & -0.969 & -0.585 & -1.056 \\ 
  I(RM\verb|^|2) & 0.007 & 0.007 & -0.006 & -0.007 & -0.002 & -0.007 \\ 
  AGE & -0.000 & -0.000 & 0.003 & 0.003 & 0.001 & 0.002 \\ 
  log(DIS) & -0.130 & -0.132 & -0.218 & -0.224 & -0.076 & -0.221 \\ 
  log(RAD) & 0.058 & 0.058 & 0.068 & 0.069 & 0.014 & 0.078 \\ 
  TAX & -0.000 & -0.000 & 0.000 & 0.000 & 0.000 & 0.000 \\ 
  PTRATIO & -0.012 & -0.012 & -0.024 & -0.024 & -0.016 & -0.027 \\ 
  B & 0.000 & 0.000 & -0.000 & -0.000 & -0.000 & -0.001 \\ 
  log(LSTAT) & -0.203 & -0.203 & -0.195 & -0.196 & -0.112 & -0.180 \\ 
   \hline
\end{tabular}
\end{scriptsize}
\caption{Indirect impacts, Boston housing data.}
\label{tab:iiboston}

\end{table}



\begin{figure}[h!]
\begin{center}
\includegraphics[width=16cm]{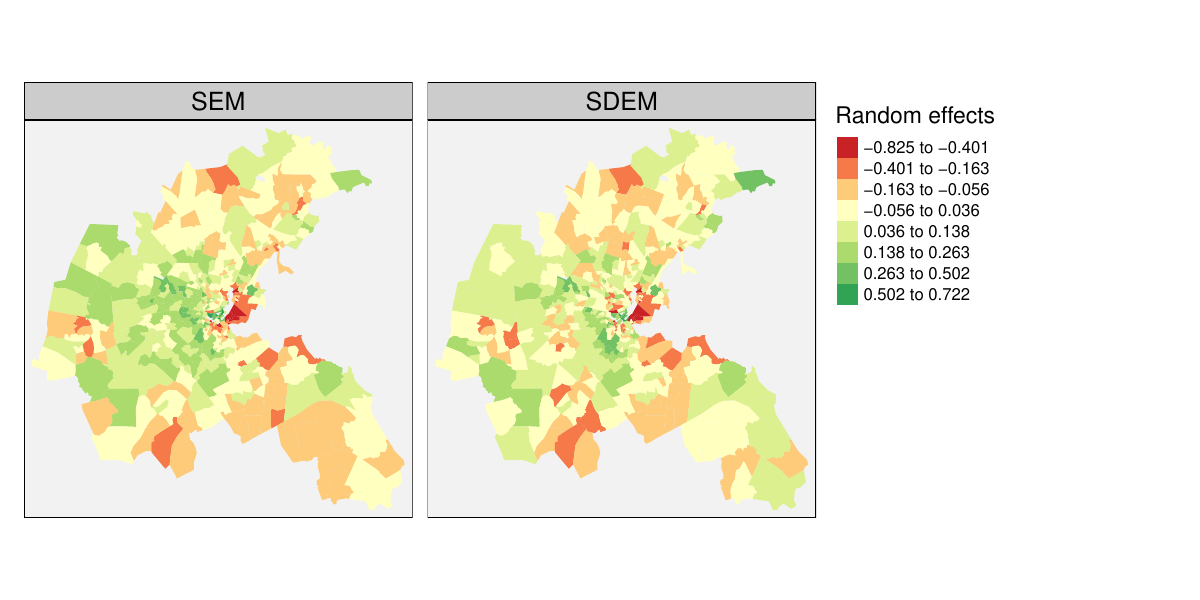}
\end{center}
\caption{Posterior means of the \code{slm} latent effects for the SEM and SDEM
models, Boston housing data with full adjacency matrix.}
\label{fig:slm-effects-full}
\end{figure}

\ 

\newpage

\

\newpage

\section*{Katrina business data: supplementary materials}

\begin{table}[h!]
\centering
\begin{tabular}{rrrrr}
  \hline
 & SEM & SLM & SDM & SDEM \\
  \hline
mean & 0.605 & 0.513 & -0.113 & 0.550 \\
  sd & 0.095 & 0.098 & 0.434 & 0.137 \\
  0.025 quantile & 0.381 & 0.290 & -1.115 & 0.209 \\
  0.975 quantile & 0.753 & 0.676 & 0.505 & 0.744 \\
   \hline
\end{tabular}

\caption{Summary of point estimates of the spatial autocorrelation parameters,
Katrina business data.}
\label{tab:Krho}
\end{table}


\begin{table}[h!]
\centering
\begin{scriptsize}
\begin{tabular}{rrrrrr}
  \hline
 & INLASEM & INLASLM & INLASDM & INLASDEM & INLASLX \\ 
  \hline
flood\_depth & -0.086 & -0.076 & -0.083 & -0.078 & -0.076 \\ 
  log\_medinc & 0.381 & 0.316 & 0.250 & 0.239 & 0.242 \\ 
  small\_size & -0.070 & -0.173 & -0.287 & -0.192 & -0.271 \\ 
  large\_size & -0.078 & -0.208 & -0.371 & -0.446 & -0.347 \\ 
  low\_status\_customers & -0.067 & -0.201 & -0.339 & -0.333 & -0.321 \\ 
  high\_status\_customers & 0.027 & 0.046 & 0.028 & 0.059 & 0.029 \\ 
  owntype\_sole\_proprietor & 0.154 & 0.338 & 0.420 & 0.336 & 0.392 \\ 
  owntype\_national\_chain & 0.016 & 0.050 & 0.406 & 0.506 & 0.353 \\ 
   \hline
\end{tabular}
\end{scriptsize}
\caption{Total impacts, Katrina data set.}
\label{tab:tikatrina}
\end{table}

\begin{table}[h!]
\centering
\begin{scriptsize}
\begin{tabular}{rrrrrr}
  \hline
 & INLASEM & INLASLM & INLASDM & INLASDEM & INLASLX \\ 
  \hline
flood\_depth & 0.000 & -0.037 & 0.018 & 0.021 & 0.014 \\ 
  log\_medinc & 0.000 & 0.157 & -0.254 & -0.169 & -0.144 \\ 
  small\_size & 0.000 & -0.088 & -0.190 & -0.113 & -0.192 \\ 
  large\_size & 0.000 & -0.108 & -0.286 & -0.366 & -0.279 \\ 
  low\_status\_customers & 0.000 & -0.101 & -0.305 & -0.316 & -0.304 \\ 
  high\_status\_customers & 0.000 & 0.024 & 0.019 & 0.045 & 0.018 \\ 
  owntype\_sole\_proprietor & 0.000 & 0.173 & 0.219 & 0.170 & 0.223 \\ 
  owntype\_national\_chain & 0.000 & 0.026 & 0.346 & 0.473 & 0.320 \\ 
   \hline
\end{tabular}
\end{scriptsize}
\caption{Indirect impacts, Katrina data set.}
\label{tab:iikatrina}
\end{table}

\end{document}